\documentclass{JHEP3}

\usepackage[latin1]{inputenc}
\usepackage[dvips,all]{xy}

\usepackage{array}
\usepackage{booktabs}
\usepackage[mathscr]{eucal}

\usepackage{amsthm}
\newtheorem{cond}{Condition}
\newtheorem{dfn}{Definition}
\newtheorem{conj}{Conjecture}
\newtheorem{thm}{Theorem}
\newtheorem{prop}[thm]{Proposition}
\newtheorem{lemma}[thm]{Lemma}

\newenvironment{twomatrix}{\biggl ( \begin{matrix}}{\end{matrix} \biggr )}

\usepackage{mysymb}
\usepackage{myref}

\newcolumntype{x}{>{$}r<{$}@{}>{$}l<{$}}

\newcommand{\Id}{\operatorname{Id}}

\newcommand{\TM}{\mathscr{T}}
\newcommand{\smpmatrix}[1]{\bigl(\begin{smallmatrix}#1%
                                 \end{smallmatrix}\bigr)}
\newcommand{\rAngle}{\rangle\!\rangle} 
\newcommand{\normord}[1]{\mathord{:\negthickspace#1\negthickspace:}}
\accentedsymbol{\Gbar}{\bar{G}}
\accentedsymbol{\TMp}{\TM'}

\newcommand{\ulab}[1]{\save[]+/u6mm/*\txt{\tiny $#1$}\restore}
\newcommand{\dlab}[1]{\save[]+/d6mm/*\txt{\tiny $#1$}\restore}

\title{Moduli spaces and D-brane categories of tori using SCFT}
\author{Christian van Enckevort\\
Johannes Gutenberg-Universität, Mainz, Germany\\
E-mail: \email{enckevor@mathematik.uni-mainz.de}}

\abstract{We analyse the moduli spaces of superconformal field
  theories (SCFTs). For $N=2$ we find an enhanced moduli space which
  in geometrical terms corresponds to tori with two independent
  complex structures.  To explain the precise relation with the moduli
  space of SCFTs on K3 surfaces as described by Aspinwall and
  Morrison, we discuss some subtleties with the precise interpretation
  of the $N=2$ and $N=4$ moduli spaces. We also explain why in some
  cases the SYZ-description of mirror symmetry as fibrewise T-duality
  seems to break down.

  Using gluing matrices we give an algebraic description of D-branes
  and construct the corresponding boundary states. We study how
  isomorphisms of the SCFTs act on D-branes. Finally we give a
  geometrical interpretation of our algebraic constructions and make
  contact with the geometrical D-brane categories and Kontsevich's
  homological mirror symmetry conjecture.}

\keywords{Mirror symmetry, vertex algebra, conformal field theory, torus}

\preprint{\hepth{0302226}}

\numberwithin{equation}{section}

\begin{document}

\section{Introduction}
Tori are by far the simplest Calabi-Yau spaces and many aspects of
string theory compactified on a torus are pretty well understood.
Nevertheless some of the more subtle aspects still deserve careful
investigation.  In this paper we study the moduli space of conformal
field theories with a torus as target space. We also investigate the
description of D-branes. The conformal field theory point of view
often leads to a rather algebraic description. We try to make the
connection with the geometry of the tori precise. Let us give a quick
overview and point out the most important results.

We start with some general remarks about the moduli space of $N=1$ and
$N=2$ superconformal field theories (SCFTs). This allows us to
introduce some notation and explain why from the SCFT point of view it
makes sense to work with two independent complex structures to
describe the $N=2$ structures corresponding to a fixed $N=1$
structure. We discuss how the $N=2$ and $N=4$ moduli spaces depend on
the precise definition of $N=2$ and $N=4$ superconformal algebras.
This leads to two slightly different moduli spaces for both $N=2$ and
$N=4$. Using this description we can clarify the connection of our
description for higher dimensional tori with the one found in the
literature for K3 surfaces and 4-dimensional tori. It also leads to
the notion of generalised morphisms of $N=2$ SCFTs, which includes
left and right mirror morphisms.

To obtain a concrete construction of the moduli space for tori we
review the description by Kapustin and Orlov in~\cite{KO00}. For $N=2$
SCFTs their description contains one not very natural condition.
Dropping this condition we find a larger moduli space containing
theirs. This enhanced moduli space turns out to be of the general form
that we discussed before. To see this we discuss a geometrical
interpretation of this larger moduli space involving a pair of complex
structures on the torus. In these terms the extra condition used by
Kapustin and Orlov has a natural interpretation: it is equivalent to
the requirement that the two complex structures coincide.

The spaces that we discuss are not really moduli spaces because there
are many isomorphisms of conformal field theories that identify points
in these spaces. Kapustin and Orlov give a concrete description of
isomorphisms and mirror morphisms of conformal field theories and
their description generalises naturally to the enhanced moduli space
we discuss. It turns out that in general the condition that the two
complex structures coincide is not invariant under isomorphisms or
mirror morphisms. This explains why for certain background field
configurations mirror symmetry seemed to fail.

We continue by studying how D-branes fit into this picture. Using the
conformal field theory description of D-branes in terms of gluing
matrices, we define a large class of D-branes. We describe the
corresponding boundary states and discuss how they transform under
generalised morphisms.  Following~\cite{OOY96}, we propose a
geometrical interpretation of the D-branes as a module over the
algebra of functions on the target space. This description is a slight
modification of the standard description of D-branes as bundles on a
submanifold.

According to the standard philosophy, D-branes can be interpreted as
the objects of the so-called D-brane categories. After a topological
twist these categories are expected to correspond to the Fukaya
category for the A-twist and the derived category of coherent sheaves
for the B-twist.  In this geometrical interpretation a mirror morphism
should induce an equivalence of the Fukaya category and the derived
category of the mirror. This is Kontsevich's famous homological mirror
symmetry conjecture. For the torus our description is more detailed
and at least in principle also works when mirror symmetry does not
preserve the condition that both complex structures are equal. Using
our description of the action of generalised morphisms on D-branes we
also obtain a fairly good understanding of the equivalence functor on
the level of objects. Making this more precise and extending it to
morphisms would lead to a much better understanding of this
conjecture.

\section{Moduli space of SCFTs}
\subsection{General structure}
\label{sec:gensit}
Before specialising to SCFTs on tori we will discuss some general
properties of SCFTs and their moduli spaces. Note that most of the
time we do not discuss the moduli spaces themselves, but talk about
the \emph{Teichmüller spaces} instead. The actual moduli space can be
obtained as the quotient of the Teichmüller space by a discrete group,
the so-called \emph{duality group}.

Our starting point is the Teichmüller space of $N=1$ SCFTs, which we
denote by $\TM_{N=1}$. For any manifold $X$ with a Ricci-flat metric
$G$ we can write down the action of a nonlinear sigma model defining a
SCFT with $N=1$ supersymmetry. The most general form of this action
includes a term depending on a closed 2-form $B$ on the target space
$X$. This 2-form is often called the B-field. Different pairs $(G,B)$
and $(G',B')$ may correspond to isomorphic $N=1$ SCFTs. Under
favourable circumstances there exist extra conditions on the metric
and the B-field that can always be fulfilled up to isomorphism. Then
we can resolve part of the ambiguity by imposing these conditions. For
the torus we can require the metric and the B-field to be
constant. This defines the Teichmüller space $\TM_{N=1}$.

However, these extra conditions will not get rid of all isomorphisms.
At least in the case of the torus the identifications due to the
remaining isomorphisms can be described by the action of a discrete
duality group $\mathcal{G}$. Therefore the moduli space can be written
as $\mathcal{M}_{N=1} = \mathcal{G}\backslash \TM_{N=1}$.

To describe the Teichmüller space for SCFTs with more supersymmetry, we
have to be more specific. First of all, note that we do not discuss
space-time supersymmetry in this text. We also ignore the possibility
of having different amounts of supersymmetry in the left- and
right-moving sectors. So when we talk about $N=1$ supersymmetry, this
means $N=1$ worldsheet supersymmetry in both the left- and the
right-moving sector. The left-moving $N=1$ superconformal algebra is
generated by two fields $L(z)$ and $G(z)$ characterised by their
operator product expansions (OPEs). These OPEs can be translated into
commutation relations for the modes. In the geometrical situation
where the SCFT corresponds to a nonlinear sigma model, these fields
are defined in terms of the metric $G$ on the target space
\begin{equation}
\label{eq:LG}
\begin{split}
  L(z) &= \frac{1}{2} \normord{G(\partial X(z), \partial X(z))} -
  \frac{1}{2} \normord{G(\psi(z), \partial \psi(z))}, \\
  G(z) &= \frac{i}{2\sqrt{2}} \normord{G(\psi(z), \partial X(z))},
\end{split}
\end{equation}
where $\normord{\phantom{x}}$ denotes operator normal ordering. In the
sigma model language the bosonic field $X(z,\bar{z})$ corresponds to
the embedding of the worldsheet into the target manifold. In itself it
is neither left-moving (holomorphic) nor right-moving
(antiholomorphic). However, the derivatives $\partial X(z)$ and
$\bar{\partial} X(\bar{z})$ are left- respectively right-moving.  The
remaining field $\psi(z)$ is fermionic and has a right-moving
counterpart $\bar{\psi}(\bar{z})$.  The expressions for the generators
$\bar{L}(\bar{z})$ and $\Gbar(\bar{z})$ in the right-moving sector can
be obtained from the ones above by replacing $\partial X(z)$ by
$\bar{\partial}X(\bar{z})$ and $\psi(z)$ by $\bar{\psi}(\bar{z})$.
Mathematically the precise interpretation of these expressions for
general Calabi-Yau manifolds is delicate, but for tori, where the
Ricci-flat metric $G$ can be represented by a constant matrix,
everything can be made rigorous (see~\cite{KO00}).

For $N=2$ superconformal algebras there are two slightly different
definitions. The first definition says that an $N=2$ superconformal
algebra consists of an $N=1$ superconformal algebra together with a
choice of a $\mathrm{u}(1)$ subalgebra with certain properties. The
second definition defines an $N=2$ superconformal algebra as an $N=1$
superconformal algebra with a choice of fields $J(z)$ and $G^\pm(z)$
satisfying certain OPEs (also involving $L(z)$).  The relation between
these two definitions is that the field $J(z)$ is a generator of the
$\mathrm{u}(1)$ subalgebra.  However, the $\mathrm{u}(1)$ subalgebra
only determines $J(z)$ up to a scalar. The required OPEs restrict this
scalar to $\pm 1$, but this remaining indeterminacy cannot be
eliminated. Once we have fixed $J(z)$, we can define $G^\pm(z)$ using
$J(z)$ and $G(z)$. The properties of the $\mathrm{u}(1)$ subalgebra
should ensure that the commutation relations are satisfied.

The Teichmüller space of $N=2$ SCFTs depends on the definition we use.
Let us write $\TM_{N=2}$ in case of the first definition and
$\TMp_{N=2}$ when we use the second definition. Then $\TMp_{N=2}$ is a
$\Z_2 \times \Z_2$-fibration over $\TM_{N=2}$. The fibres correspond
to the choice of the signs of $J(z)$ and the corresponding field
$\bar{J}(\bar{z})$ in the right-moving sector. In this case the
difference between the two Teichmüller spaces is rather small, but
conceptually it is quite important.  As we will see in the next
section for $N=4$ a similar phenomenon occurs which will be important
to explain the relation between our description of the moduli space
and the one in the literature.

Isomorphisms of $N=2$ SCFTs can be defined as $N=1$ isomorphisms that
preserve the $\mathrm{u}(1)$ algebras for the first definition or the
generators for the second one. In both cases the duality group can be
identified with the duality group $\mathcal{G}$ for the $N=1$ case.
The reason is that any $N=2$ isomorphism is also an $N=1$ isomorphism.
Conversely, every $N=1$ isomorphism determines an $N=2$ isomorphism. So
the moduli spaces are given by $\mathcal{M}_{N=2} = \mathcal{G}
\backslash \TM_{N=2}$ and $\mathcal{M}'_{N=2} = \mathcal{G}
\backslash \TM'_{N=2}$ respectively.

In the sequel we will follow Kapustin and Orlov (see~\cite{KO00}) and
use the second definition. It may be less elegant than the first one,
but it has the significant advantage that the OPEs for $J(z)$ and
$G^\pm(z)$ are well known, whereas the precise conditions for the
$\mathrm{u}(1)$ subalgebra are unclear.  A consequence of using this
definition is that mirror symmetry does not define an isomorphism of
$N=2$ SCFTs, as it would do if we used the first definition. Instead,
we can define \emph{left mirror morphisms} as $N=1$ isomorphisms that
in the left-moving sector change the sign of $J(z)$ and interchange
$G^\pm(z)$, whereas in the right-moving sector they preserve
$\bar{J}(\bar{z})$ and $\bar{G}^\pm(\bar{z})$.  Similarly, \emph{right
  mirror morphisms} change the sign of right-moving current
$\bar{J}(\bar{z})$ and preserve $J(z)$. Together, left and right
mirror morphisms generate a group $\mathcal{G}' = \mathcal{G} \times
\Z_2 \times \Z_2$, which we will call the \emph{extended duality
  group}. Dividing out this group we obtain the moduli space
corresponding to the first definition $\mathcal{M}_{N=2} =
\mathcal{G}' \backslash \TM'_{N=2}$.

Forgetting the $N=2$ structure defines maps $\TM_{N=2} \rightarrow
\TM_{N=1}$ and $\TMp_{N=2} \rightarrow \TM_{N=1}$. In both cases the
fibres of these maps can be written as $S \times S$, where $S$ is the
space of all $N=2$ structures for a fixed $N=1$ structure. Of course
the space $S$ differs depending on which definition we use. The two
factors correspond to the left- and right-moving sectors.

For the geometrical interpretation it is also advantageous to use the
second definition. For a nonlinear sigma model to allow $N=2$
supersymmetry the target space should be a Kähler manifold. Supposing
that $X$ is a Kähler manifold, we can write down explicit formulae for
the fields in terms of geometrical data
\begin{equation}
\label{eq:Ntwogens}
\begin{aligned}
  G^{\pm}(z) &= \frac{i}{4\sqrt{2}} \normord{G(\psi(z), \partial
    X(z))} \pm \frac{1}{4\sqrt{2}}
  \normord{\omega(\psi(z), \partial X(z))}, \\
  J(z) &= - \frac{i}{2} \normord{\omega(\psi(z),\psi(z))}.
\end{aligned}
\end{equation}
Here $\omega$ is the Kähler form on the target space. Finding a Kähler
form is equivalent to finding a complex structure $j$ compatible with
the metric $G$ (i.e., satisfying $j^t G j = G$). The expressions for
the right-moving sector can again be obtained by putting bars in the
appropriate places. However, note that we could use a different
complex structure (but the same metric) in the right-moving sector. So
we can choose two independent complex structures $j_1$ and $j_2$ both
compatible with the same metric $G$ and use $j_1$ to define $J(z)$ and
$G^\pm(z)$ and $j_2$ to define $\bar{J}(\bar{z})$ and
$\bar{G}^\pm(\bar{z})$. All we have to check is that these fields
satisfy the required OPEs. For this it does not matter whether $j_1 =
j_2$ or not: the OPEs within a sector are independent of the complex
structure in the other sector and operators from different sectors
should (anti)commute and that is independent of the complex structure.

This means that for the second definition we can interpret $S$
geometrically as the space of complex structures compatible with a
given metric. Therefore the real dimension of $S$ is $2h^{2,0}$, where
$h^{p,q} = h^{p,q}(X)$ are the Betti numbers of the target space $X$.
In fact it seems that higher supersymmetries occur exactly when
$h^{2,0} > 0$, i.e., when $S$ has a positive dimension.  This
situation can be summarised in the following diagram (with the real
dimension above or below each space)
\begin{equation}
\label{eq:Ntwodiag}
\xymatrix{
  \TMp_{N=2}^{\mathrm{geom}} \ulab{D+2h^{2,0}}
  \ar@{^{(}->}[r]_{\mathrm{diag.}} \ar[dr] &
  \TMp_{N=2} \ulab{D+4h^{2,0}} \ar[d] & 
    S \times S \ulab{2 \times 2h^{2,0}} \ar[l] \\
  & \TM_{N=1} \dlab{D}}
\end{equation}
Here $n=\dim_\C X$ and $D=2(h^{1,1} + h^{n-1,1})$ is the real dimension
of $\TM_{N=1}$.  This dimension can be computed as follows.
Deformations of the complex structure contribute $2h^{n-1,1}$, the
choice of the B-field $b_2 = h^{1,1} + 2 h^{2,0}$ and the choice of
the Kähler form $h^{1,1}$.  Because for $N=1$ the complex structure is
irrelevant (only the metric and the B-field count), one should
subtract $2h^{2,0}$ for the choice of a complex structure for one
fixed metric.

The geometrical part $\TMp_{N=2}^{\mathrm{geom}}$ is where the
complex structures in both sectors are equal. So fibrewise it is the
diagonal in $\TM'_{N=2}$.  A slightly degenerate case of this picture
is obtained when the target space is a strict Calabi-Yau manifold
(i.e., a manifold with holonomy group equal to $\mathrm{SU}(n)$ for
$n>2$). In that case there is just a $\Z_2$ of choices for the complex
structure, so $S = \Z_2$. This also means that $\TM_{N=2}$ is equal
to $\TM_{N=1}$. For strict Calabi-Yau manifolds $N=2$ is the maximal
amount of supersymmetry. It seems to be a general phenomenon that if
one considers a SCFT on a certain target space with the maximal amount
of supersymmetry $N=N_\mathrm{max}$ possible for that target space,
then $\TM_{N=N_\mathrm{max}} = \TM_{N=1}$. Maximal supersymmetry
should correspond to choosing the largest possible subalgebra from
some series of subalgebras indexed by $N$. Then the conjecture is that
for a given $N=1$ SCFT there is just one possible choice for this
maximal subalgebra. We will see another example of this in the next
section.

In general mirror symmetry does not preserve
$\TMp_{N=2}^{\mathrm{geom}}$. A trivial example is the (left) mirror
morphism that is the identity on the $N=1$ level, but replaces $J(z)$
by $-J(z)$ and interchanges $G^+(z)$ and $G^-(z)$. Geometrically this
corresponds to replacing $j_1$ by $-j_1$. Of course, this is a rather
trivial mirror morphism. The geometrically interesting mirror
morphisms do preserve $\TM_{N=2}^{\mathrm{geom}}$.

\subsection{K3 surfaces and 4-tori}
Apart from the degenerate case of strict Calabi-Yau manifolds the
prime example in the literature of the structure discussed in the
previous section is the moduli space of SCFTs on K3 surfaces and
4-tori. Their moduli spaces are very similar and have been studied in
great detail (see e.g.,~\cite{AM94,Dij98,NW99,NW01}). In those papers
most attention was paid to the moduli space of $N=4$ SCFTs. In the
previous section we encountered some subtleties in the interpretation
of $N=2$ moduli spaces. For $N=4$ superconformal algebras there are
also two slightly different definitions. One possibility is to define
an $N=4$ superconformal algebra as an $N=1$ superconformal algebra
with an $\mathrm{su}(2)$ subalgebra.  The other possibility is to
choose generators and define an $N=4$ algebra as an $N=1$ algebra with
extra fields $J^{(i)}(z)$ ($i=1,2,3$), $G^\pm(z)$, and ${G'}^\pm(z)$
satisfying certain OPEs. The corresponding Teichmüller spaces are
denoted by $\TM_{N=4}$ and $\TM'_{N=4}$ respectively.

Recall that K3 surfaces and 4-tori are hyperkähler manifolds. In
particular for a fixed metric there exists an $S^2$ of compatible
complex structures. More explicitly, we can choose three compatible
complex structure $j_i$ ($i=1,2,3$) satisfying the quaternion
relations $j_1 j_2 = j_3$ and cyclic permutations. An arbitrary
compatible complex structure $j$ can then be written as $j = \sum_k
a_k j_k$ with $(a_1, a_2, a_3) \in S^2 \subset \R^3$. If denote the
corresponding Kähler forms by $\omega_i$, then we can define the
generators $J^{(i)}(z)$ of the $\mathrm{su}(2)$ as
\[
  J^{(i)}(z) = - \frac{i}{2} \normord{\omega_i(\psi(z),\psi(z))}.
\]
These fields generate the $\mathrm{su}(2)$ subalgebra. Note that this
subalgebra only depends on the metric. It follows that there is just
one choice for the $\mathrm{su}(2)$ subalgebra and therefore
$\TM_{N=4} \cong \TM_{N=1}$.

For further reference we will briefly sketch the known description of
the moduli space for K3 surfaces and 4-tori. Using the uniform
notation introduced in~\cite{NW99} we can discuss both cases
simultaneously. Let $X$ denote a K3 surface or a 4-torus, then we
define $\Lambda$ to be the lattice $\Hgrp^{\mathrm{ev}}(X, \Z) =
\oplus_k \Hgrp^{2k}(X, \Z)$ of even integral cohomology classes. On
this lattice there exists an even pairing $q$ defined by $q(\alpha,
\beta) = -\int_X \alpha \wedge I(\beta)$, which is a slight
modification of the intersection pairing.  In this formula $I:
\Hgrp^{\mathrm{ev}}(X, \Z) \rightarrow \Hgrp^{\mathrm{ev}}(X, \Z)$ is
defined by $I|_{\Hgrp^{2k}(X, \Z)} = (-1)^k \Id$. The signature of
this pairing is $(4, 4+\delta)$, where $\delta = 0$ for a 4-torus and
$\delta = 16$ for a K3 surface.

A SCFT with target space a K3 surface or a 4-torus has $N=4$
worldsheet supersymmetry. The $N=4$ Teichmüller space is given by
\begin{equation}
\label{eq:TMNfour}
  \TM_{N=4}(X) = O(\Lambda_\R,q)/O(4) \times O(4+\delta),
\end{equation}
where $\Lambda_\R := \Lambda \otimes \R = \Hgrp^{\mathrm{ev}}(X, \R)$.
The duality group turns out to be $O(\Lambda,q)$, the group of all
lattice automorphisms of $\Lambda$ preserving $q$. For K3 surfaces
this was originally a conjecture by Seiberg (see~\cite{Sei88}).
In~\cite{AM94} Aspinwall and Morrison analysed the duality group in
great detail and argued that Seiberg's guess is indeed correct. The
remaining issues were settled by Nahm and Wendland in~\cite{NW99}.
Using the duality group we find the following description of the
$N=4$ moduli space
\begin{equation}
\label{eq:MNfour}
  \mathcal{M}_{N=4}(X) = O(\Lambda,q) \backslash O(\Lambda_\R,q)/O(4) 
    \times O(4+\delta).
\end{equation}
Aspinwall and Morrison describe the $N=2$ Teichmüller space for K3
surfaces as a fibration $\TMp_{N=2} \rightarrow \TM_{N=4}$ with fibres
$S^2 \times S^2$. This coincides with the description that we found in
the previous section when we use that $\TM_{N=4} = \TM_{N=1}$. Note
that Aspinwall and Morrison do choose generators $J(z)$ and
$\bar{J}(\bar{z})$ for the $N=2$ SCFT\@, but they do not do this for
the $N=4$ SCFT\@. From our perspective, the existence of a map
$\TMp_{N=2} \rightarrow \TM_{N=4}$ is a coincidence ultimately
stemming from the fact that for K3 surfaces $\TM_{N=4}$ is equal to
$\TM_{N=1}$ and there is a natural map $\TMp_{N=2} \rightarrow
\TM_{N=1}$.

\subsection{$N=1$ SCFTs on tori}
After these generalities we will restrict our attention to tori for
the rest of this paper. To give a precise description of the moduli
spaces, we will use the description of SCFTs for tori due to Kapustin
and Orlov (see~\cite{KO00}). Note that they mostly discuss vertex
algebras, which provide a mathematical formulation of most ingredients
that go into a (S)CFT\@.  Because the vertex algebras for tori that they
describe can be extended to full (S)CFTs, we will be sloppy and use
the names vertex algebra and SCFT interchangeably.  Many of the
formulae we discuss in this section are known in the physics
literature (see e.g.,~\cite{GPR94}), but we want to collect them here
to establish the notation and prepare the ground for the discussion of
the $N=2$ case which is less well known.

Let $X$ be a torus of real dimension $m$ equipped with a Ricci-flat
metric $G$ and a closed $2$-form $B$. Let $\Gamma$ be the lattice
$\Hgrp_1(X, \Z)$. We will write $\Gamma_\R$ for the corresponding real
vector space $\Gamma \otimes \R$, so we can write $X =
\Gamma_\R/\Gamma$. The Ricci-flat metric and the B-field can then be
represented by linear maps $\Gamma_\R \to \Gamma_\R^*$, which we will
also denote by $G$ and $B$. Note that $G$ is symmetric, whereas $B$ is
antisymmetric.

Using these data Kapustin and Orlov define an $N=1$ SCFT for every
torus. We will briefly discuss some essential points of their
construction below, but first we want to present a different way of
encoding the metric and the B-field. To do so, let us first introduce
some notation. On the lattice $\Gamma \oplus \Gamma^*$ we have a
symmetric bilinear form $q$ defined by $q(v \oplus f, w \oplus g) =
f(w) + g(v)$. When we talk about a bilinear form $q$ in connection
with a lattice of the form $\Gamma \oplus \Gamma^*$ we will always
mean this particular bilinear form. A lattice isomorphism $g: (\Gamma
\oplus \Gamma^*,q) \rightarrow (\Gamma' \oplus {\Gamma'}^*,q')$ will
be supposed to respect the bilinear forms in the sense that $g^t q' g
= q$.

In terms of the metric $G$ and the B-field $B$ we can define an
endomorphism $K$ of $\Gamma_\R \oplus \Gamma_\R^*$ by
\begin{equation}
\label{eq:K}
  K = \begin{twomatrix}
          -G^{-1}B   &  G^{-1} \\
        G - BG^{-1}B & BG^{-1}
      \end{twomatrix}.
\end{equation}
One can easily check that $K^2 = \Id$ and $K^t q K = q$. To ensure
that conversely any such $K$ can be written in the above form, we need
one more requirement. This comes from the condition that $G$ be
positive definite. To investigate this condition, let us define
\[
  \mathcal{R}(G,B) 
   = \frac{1}{2} \begin{twomatrix}
                   \Id_m - G^{-1}B & \phantom{-}G^{-1} \\
                   \Id_m + G^{-1}B & -G^{-1}
                 \end{twomatrix}.
\]
This is a slight modification of the definition in~\cite{KO00}. A
straightforward computation shows that
\begin{equation}
\label{eq:qKdiag}
  qK = \begin{twomatrix}
         G - BG^{-1}B & BG^{-1} \\
           -G^{-1}B   &  G^{-1}  
       \end{twomatrix}
     = \mathcal{R}(G,B)^t \begin{twomatrix} 2G & 0 \\ 0 & 2G \end{twomatrix}
       \mathcal{R}(G,B).
\end{equation}
So $G$ is positive definite if and only if $qK$ is. Putting
all these ingredients together we are led to the following definition
of the $N=1$ Teichmüller space
\begin{dfn}
\label{def:MNone}
The Teichmüller space of $N=1$ SCFTs on a torus of real dimension $m$ is
defined as
\[
  \TM_{N=1} := \biggl \{ (\Gamma,K) \biggm | \negthickspace
  \begin{array}{l}
    \text{$\Gamma$ lattice of rank $m$,
          $K \in \End(\Gamma_\R \oplus \Gamma^*_\R)$}\\
    \text{$K^2 = \Id$, $K^t q K = q$, $q K$ pos.\ def.}
  \end{array} \biggr \}.
\]
\end{dfn}
Here $q$ is the symmetric bilinear form on the lattice $\Gamma \oplus
\Gamma$ discussed above.  This space has several well known
alternative descriptions which we summarise in the following
proposition. For the sake of completeness we also sketch a proof.
\begin{prop}
\label{prop:MNone}
The $N=1$ Teichmüller as defined in \sref{def:MNone} has the following
alternative descriptions
\begin{align*}
  \TM_{N=1} &= \biggl \{ (\Gamma,G,B) \biggm | \negthickspace
  \begin{array}{l}
    \text{$\Gamma$ lattice of rank $m$, $B,G: \Gamma_\R \to
      \Gamma_\R^*$,} \\
    \text{$G$ symm., $B$ antisymm., $G$ pos.\ def.}
  \end{array} \biggr \}\\
  &= \{ (\Gamma,E) \mid \text{$\Gamma$ lattice of rank $m$,
   $E: \Gamma_\R \to \Gamma_\R^*$, $E + E^t$ pos.\ def.} \}\\
  &= \{ (\Gamma,V) \mid \text{$V \subset \Gamma_\R \oplus \Gamma^*_\R$,
  $V$ maximal pos.\ subspace w.r.t. $q$} \}.
\end{align*}
\end{prop}
\begin{proof}
  For the first equality, note that the inclusion $\supset$ follows
  from \pref{eq:K}. To prove the other inclusion we need to show that
  any $K \in \TM_{N=1}(\Gamma)$ can be written in the form
  \pref{eq:K} with $G$ symmetric and positive definite and $B$
  antisymmetric. To see this write $K = \smpmatrix{\alpha & \beta \\ 
    \gamma & \delta}$.  Now note that $\beta: \Gamma_\R^* \rightarrow
  \Gamma_\R$ has to be invertible, because otherwise $\langle v, q K v
  \rangle = 0$ for $v \in 0 \oplus \Gamma_\R^*$ and that contradicts
  the requirement that $q K$ be positive definite.  Knowing this, one
  can easily show that $K^2 = \Id$ is equivalent to $\gamma =
  \beta^{-1} - \beta^{-1} \alpha^2$ and $\delta = -\beta^{-1} \alpha
  \beta$. Substituting that into $K^t q K = q$, one finds that
  $\beta^{-1}\alpha$ is antisymmetric and $\beta$ is symmetric. So if
  we identify $\beta$ with $G^{-1}$ and $\beta^{-1}\alpha$ with $-B$,
  then $K$ will have the form \pref{eq:K}.  The positive definiteness
  of $G$ follows from the positive definiteness of $qK$ using
  \pref{eq:qKdiag}.
  
  The second alternative is just a reformulation of the first one: $E$
  can be defined as $B+G$ and $B$ and $G$ can be obtained from $E$ as
  the symmetric and antisymmetric part.
  
  To show the third equality note that $K^2 = \Id$ implies that $K$
  is diagonalisable with eigenvalues $\pm 1$.  Let $E_{\pm 1}$ be the
  eigenspaces of $K$ for the eigenvalues $\pm 1$.  Then for all $v \in
  E_{+1} \setminus \{ 0 \}$ we have
\[
  0 < \langle v, qK v \rangle = \langle v, q v \rangle.
\]
So $q$ restricted to $E_{+1}$ is positive definite. Similarly, it
follows that $q$ restricted to $E_{-1}$ is negative definite. This
means that $E_{\pm1}$ can each have at most dimension $m$. Because
$E_{+1} \oplus E_{-1} = \Gamma_\R \oplus \Gamma_\R^*$, we can conclude
that $E_{\pm 1}$ indeed have dimension $m$. So $E_{+1}$ is a maximal
positive subspace. Conversely, if $V$ is a maximal positive subspace
for $q$, then $W := V^\perp$ is a maximal negative subspace. So we can
define an endomorphism $K$ of $\Gamma_\R \oplus \Gamma_\R^* = V \oplus
W$ that is the identity on $V$ and minus the identity on $W$. One can
easily check that $K$ defined in this way satisfies the conditions
stated above in the definition of $\TM_{N=1}$.

Alternatively, note that if $E$ is a map $\Gamma_\R \to \Gamma_\R^*$
with positive definite symmetric part, then $\phi_E = \smpmatrix{\Id
  \\ E}$ defines a map $\Gamma_\R \to \Gamma_\R \oplus \Gamma_\R^*$.
One can easily check that the image $V = \im \phi_E$ is a positive
subspace and that $V^\perp = \im \phi_{-E^t}$.
\end{proof}

Because any two lattices of the same rank are isomorphic, it often
makes sense to make one fixed choice of the lattice. So we could
define an alternative Teichmüller space as follows
\[
  \TM_{N=1}(\Gamma) := \{ K \mid (\Gamma,K) \in \TM_{N=1} \}.
\]
The statement of \sref{prop:MNone} continues to hold with the lattice
$\Gamma$ fixed throughout. In fact we can even extend the list with
following rather well known description of the Teichmüller space
\begin{equation}
\label{eq:MoneO}
  \TM_{N=1}(\Gamma) = O(\Gamma_\R \oplus \Gamma^*_\R,q)/O(m) \times
  O(m).
\end{equation}
Here the $O(\Gamma_\R \oplus \Gamma^*_\R,q)$ is group of automorphisms
of the lattice $\Gamma \oplus \Gamma^*$ preserving the bilinear form
$q$. This equality is based on the fact that the group $O(\Gamma_\R
\oplus \Gamma^*_\R,q)$ acts transitively on the space of maximal
positive subspaces of $(\Gamma_\R \oplus \Gamma^*_\R,q)$. Let $V$ be
any such subspace, then $\Gamma_\R \oplus \Gamma^*_\R = V \oplus
V^\perp$. The stabiliser of $V$ is $O(V) \times O(V^\perp)$. This
proves the desired equality. For $m=4$ this description coincides with
the one from \pref{eq:MNfour}.  In the sequel we will usually work with
$\TM_{N=1}$ instead of $\TM_{N=1}(\Gamma)$, because this often leads
to nicely coordinate invariant expressions.

For further reference we compute
\begin{equation}
\label{eq:Rinv}
  \mathcal{R}(G,B)^{-1} = \begin{twomatrix}
                            \Id_m & \Id_m \\
                             B+G  &  B-G
                          \end{twomatrix}
  = \begin{twomatrix}
      \Id_m & \Id_m \\
         E  &  -E^t
    \end{twomatrix}
\end{equation}
and note that this is precisely the linear transformation
diagonalising $K$, so that
\begin{equation}
\label{eq:Kdiag}
  K = \mathcal{R}(G,B)^{-1} \begin{twomatrix}
                              \Id_m &    0 \\
                                0   & -\Id_m
                            \end{twomatrix} \mathcal{R}(G,B).
\end{equation}
In addition it also block diagonalises $q$ as a bilinear form, i.e.,
\begin{equation}
\label{eq:qdiag}
  q = \mathcal{R}(G,B)^t \begin{twomatrix}
                           2G &  0 \\
                           0  & -2G
                         \end{twomatrix} \mathcal{R}(G,B).
\end{equation}
This shows again that $qK$ can be written as in
\pref{eq:qKdiag}.

So far we have just claimed that there exists an $N=1$ vertex algebra
for any $(\Gamma,K) \in \TM_{N=1}$. Let us now be slightly more
specific. We will denote the underlying (super) vector space by
$V_{\Gamma,K}$. This space is also called the state space. Sometimes
we will abuse this notation to denote the vertex algebra.
In~\cite{KO00} Kapustin and Orlov describe in detail the construction
of the state space together with its vertex algebra structure. Here we
just review some of the essentials that we will need in the sequel.

Recall that a field is an $\End(V)$-valued series in $z$ and
$\bar{z}$. More explicitly we have the following \emph{mode
  expansions} for the basic fields
\begin{equation}
\label{eq:fieldmodes}
\begin{aligned}
  \partial X(z) &= \sum_{s \in \Z} \alpha_s z^{-s-1}, \qquad & 
  \psi(z) &= \sum_{r \in \Z + \frac{1}{2}} \psi_r z^{-r-\frac{1}{2}}, \\
  \bar{\partial} X(\bar{z}) &= \sum_{s \in \Z} 
    \bar{\alpha}_s \bar{z}^{-s-1}, & 
  \bar{\psi}(\bar{z}) &= \sum_{r \in \Z + \frac{1}{2}} 
    \bar{\psi}_r \bar{z}^{-r-\frac{1}{2}}.
\end{aligned}
\end{equation}
Here $\alpha_s$, $\bar{\alpha}_s$, $\psi_r$, and $\bar{\psi}_r$ are
vectors of endomorphisms of the state space $V$ satisfying the
following (anti)commutations relations
\[
  [ \alpha_s^\mu, \alpha_r^\nu ] = s G^{\mu\nu} \delta_{s,-r}, \quad\qquad 
  \{ \psi_s^\mu, \psi_r^\nu \} = G^{\mu\nu} \delta_{s,-r},
\]
and similar relations for $\bar{\alpha}_s$ and $\bar{\psi}_r$. These
commutation relations allow us to construct the state space using the
standard Fock space construction by declaring the negative modes to be
creation operators and the positive modes to be annihilation
operators. To deal with $\alpha_0$ and $\bar{\alpha}_0$ one has to be
slightly more careful. In the end we obtain the following description
of the state space
\[
  V = V_{\Gamma,K} = \bigoplus_{(w,m) \in \Gamma \oplus \Gamma^*} V_{w,m}.
\]
Here each of the $V_{w,m}$ is a Fock space generated from a base
state $\lvert w,m \rangle$ by acting on it with the negative modes. The
\emph{sectors} $V_{w,m}$ are joint eigenspaces of $\alpha_0$ and
$\bar{\alpha}_0$ with eigenvalues $G^{-1} k_L$ and $G^{-1} k_R$, where
\begin{equation}
\label{eq:kLR}
\begin{split}
  \begin{twomatrix} k_L \\ k_R \end{twomatrix}
  = \begin{twomatrix}
       Gw - Bw + m \\
      -Gw - Bw + m
    \end{twomatrix}  
  = 2 \begin{twomatrix} G & 0 \\ 0 & -G \end{twomatrix}
  \mathcal{R}(G,B) \begin{twomatrix} w \\ m \end{twomatrix}.
\end{split}
\end{equation}
The lattice vectors $w$ and $m$ labelling the sectors are often
referred to as \emph{winding number} and \emph{momentum}. As a
side remark, note that using \pref{eq:qdiag} and \pref{eq:qKdiag} it
follows easily that
\[
\begin{split}
  \tfrac{1}{2} (k_L^2 - k_R^2)
  &:= \tfrac{1}{2} \bigl ( \begin{matrix} k_L^t & k_R^t \end{matrix} \bigr )
    \begin{twomatrix} 
      G^{-1} &    0 \\ 
         0   & -G^{-1}
    \end{twomatrix} 
    \begin{twomatrix} k_L \\ k_R \end{twomatrix}
    = \bigl ( \begin{matrix} w^t & m^t \end{matrix} \bigr )
      q \begin{twomatrix} w \\ m \end{twomatrix}, \\
  \tfrac{1}{2} (k_L^2 + k_R^2)
  &:= \tfrac{1}{2} \bigl ( \begin{matrix} k_L^t & k_R^t \end{matrix} \bigr )
    \begin{twomatrix} 
      G^{-1} &   0 \\ 
        0    & G^{-1}
    \end{twomatrix} 
    \begin{twomatrix} k_L \\ k_R \end{twomatrix}
    = \bigl ( \begin{matrix} w^t & m^t \end{matrix} \bigr )
      qK \begin{twomatrix} w \\ m \end{twomatrix}
\end{split}
\]
These formulae are well known in physics. Let us define the
\emph{charge lattice} as the lattice of all $(k_L,k_R) \in \Gamma^*_\R
\oplus \Gamma^*_\R$ that can be written in the form \pref{eq:kLR} for
$(w,m) \in \Gamma \oplus \Gamma^*$ (see~\cite{Huy01}). Then the first
formula above means that the charge lattice with a bilinear form given
by the symmetric matrix $\diag(G^{-1}, -G^{-1})$ is isomorphic to the
lattice $(\Gamma \oplus \Gamma^*,q)$.

An isomorphism of vertex algebras can be described as an isomorphism
of the underlying vector spaces preserving the vertex algebra
structure. Kapustin and Orlov give the following classification of
isomorphisms of $N=1$ vertex algebras.
\begin{thm}[Kapustin, Orlov]
\label{thm:isoNone}
  Isomorphisms between the $N=1$ SCFTs $V_{\Gamma,K}$ and
  $V_{\Gamma',K'}$ correspond 1-1 to lattice isomorphisms $g: (\Gamma
  \oplus \Gamma^*,q) \rightarrow (\Gamma' \oplus {\Gamma'}^*,q')$ satisfying
  $K' = g K g^{-1}$.
\end{thm}
Let us write $f_g$ for the isomorphism of vertex algebras
$V_{\Gamma,K} \rightarrow V_{\Gamma',K'}$ corresponding to a lattice
isomorphism $g: (\Gamma \oplus \Gamma^*,q) \rightarrow (\Gamma' \oplus
{\Gamma'}^*, q')$. This map is defined by
\begin{equation}
\label{eq:fgwm}
  f_g \lvert w,m \rangle = \lvert g \smpmatrix{w \\ m} \rangle
\end{equation}
and
\begin{equation}
\label{eq:fggens}
\begin{split}
  \begin{twomatrix} \alpha_s' f_g \\ \bar{\alpha}_s' f_g \end{twomatrix}
  &= \begin{twomatrix} 
       (a + b E) f_g \alpha_s \\ 
       (a - b E^t) f_g \bar{\alpha}_s
     \end{twomatrix}
   = M(g,E) \begin{twomatrix} 
              f_g \alpha_s \\ 
              f_g \bar{\alpha}_s
            \end{twomatrix}, \\
  \begin{twomatrix} \psi_s' f_g \\ \bar{\psi}_s' f_g \end{twomatrix}
  &= \begin{twomatrix}
       (a + b E) f_g \psi_s \\
       (a - b E^t) f_g \bar{\psi}_s
     \end{twomatrix}
   = M(g, E) \begin{twomatrix} f_g \psi_s \\ f_g \bar{\psi}_s \end{twomatrix}.
\end{split}
\end{equation}
Here $a: \Gamma \rightarrow \Gamma'$ and $b: \Gamma^* \rightarrow
\Gamma'$ are defined by writing the lattice isomorphism $g: (\Gamma
\oplus \Gamma^*, q) \rightarrow (\Gamma' \oplus {\Gamma'}^*, q')$ in
block form as $g = \smpmatrix{a & b \\ c & d}$. The matrix $E=G+B$ is
the background matrix introduced above and $M(g, E)$ is the block
diagonal matrix $M(g, E) = \diag(a + b E, a - b E^t)$. With
\pref{eq:fieldmodes} this leads to transformation formulae for the
fields (see~\pref{tab:trules}, where we will collect the various
transformation formulae). Note that compared with~\cite{KO00} we
interchanged the transformation formulae for the left- and
right-moving sectors. This is purely a matter of convention.

Using the precise definition of $f_g$ in~\cite{KO00} it is
straightforward to check the following \emph{functoriality property}.
\begin{prop}
\label{prop:Nonemorfunc}
Let $g_i: (\Gamma_i \oplus \Gamma_i^*,q_i) \rightarrow (\Gamma_{i+1}
\oplus \Gamma_{i+1}^*,q_{i+1})$ be lattice isomorphisms, then we have
the following functoriality property
\[
  f_{g_1} \circ f_{g_2} = f_{g_1 \circ g_2}.  
\]
\end{prop}
We can use \sref{thm:isoNone} to associate a map $\mu_g$ between
Teichmüller spaces to any lattice isomorphism $g$ preserving the
bilinear form. This map is defined by the property that $g$ defines an
isomorphism between the $N=1$ SCFTs $V_{\Gamma,K}$ and
$V_{\Gamma',\mu_g(K)}$.
\begin{dfn}
  Let $g: (\Gamma \oplus \Gamma^*,q) \rightarrow (\Gamma' \oplus
  {\Gamma'}^*, q')$ be a lattice automorphism, then we define the map
  $\mu_g: \TM_{N=1}(\Gamma) \rightarrow \TM_{N=1}(\Gamma')$ by
\[
  \mu_g(K) = g K g^{-1}.
\]
\end{dfn}
It is interesting to try and find a description of $\mu_g$ for the
other descriptions of the moduli space from \sref{prop:MNone}. Let us
start with the description in terms of the background field $E=G+B$.
As above we write the lattice isomorphism $g: \Gamma \oplus \Gamma^*
\rightarrow \Gamma' \oplus {\Gamma'}^*$ as $g = \smpmatrix{a & b \\ c
  & d}$, where $a: \Gamma \rightarrow \Gamma'$, $b: \Gamma^*
\rightarrow \Gamma'$ etc. Using \pref{eq:Kdiag} we find
\begin{equation}
\label{eq:Kp}
  K' = g K g^{-1} = g \mathcal{R}(G,B)^{-1} 
    \begin{twomatrix}
      \Id_m &   0 \\
        0   & -\Id_m
    \end{twomatrix}
    {(g \mathcal{R}(G,B)^{-1})}^{-1}.
\end{equation}
From \pref{eq:Rinv} it follows
\begin{equation}
\label{eq:gR}
\begin{split}
  g \mathcal{R}(G,B)^{-1} &= 
  \begin{twomatrix}
    a + b E & a - b E^t \\
    c + d E & c - d E^t
  \end{twomatrix}
  = \begin{twomatrix}
      \Id_m &  \Id_m \\
        E'  & -{E'}^t
    \end{twomatrix}
    \begin{twomatrix}
      a + b E &     0 \\
         0    & a - b E^t
    \end{twomatrix} \\
  &= \mathcal{R}(G',B')^{-1} M(g,E).
\end{split}
\end{equation}
For the second equality we defined $E' = (c + d E) (a + b E)^{-1}$.
Recall that the lattice isomorphism takes $q$ to $q'$. Using the block
form of $g$ the corresponding equation $g^t q' g = q$ can be written
as
\[
  \begin{twomatrix} a^t & c^t \\ b^t & d^t \end{twomatrix}
  \begin{twomatrix} 0 & \Id \\ \Id & 0 \end{twomatrix}
  \begin{twomatrix} a & b \\ c & d \end{twomatrix}
  = \begin{twomatrix} 0 & \Id \\ \Id & 0 \end{twomatrix}.
\]
This yields equations for $a$, $b$, $c$ and $d$ that can be used to
show that $-{E'}^t = (c - d E^t) (a - b E^t)^{-1}$. This proves the
second equality. The third equality is a matter of definition. If we
define $G'$ and $B'$ as the symmetric and antisymmetric part of $E'$,
then the first factor is equal to $R(G',B')^{-1}$. The second factor
we defined as $M(g,E)$ above. Combining \pref{eq:Kp} and \pref{eq:gR}
one sees that $K'$ can again be written in the form \pref{eq:Kdiag}
with $G'$ and $B'$ as defined above. This gives us an explicit
transformation rule for $E$ and an indirect one for $G$ and $B$. For
$G'$ one can obtain two different simplified expression starting from
the expressions for $E'$ and $-{E'}^t$ (see \pref{tab:trules}).

Because $K'$ is equal to $g K g^{-1}$, the eigenspaces of $K$
transform with $g$. So for the description of the Teichmüller space in
terms of maximal positive subspaces of $\Gamma_\R \oplus \Gamma_\R^*$
we just use $g$ to transform the subspace.

To describe the moduli space we choose a fixed target space $X$ or
equivalently a fixed lattice $\Gamma$. The duality group is given by
$\mathcal{G}= O(\Gamma \oplus \Gamma^*,q)$. So using the description
\pref{eq:MoneO} of the moduli space $\TM_{N=1}(\Gamma)$ we find the
following description of the actual moduli space
\[
  \mathcal{M}_{N=1}(\Gamma) = \mathcal{G} \backslash \TM_{N=1}(\Gamma) 
  = O(\Gamma \oplus \Gamma^*,q) \backslash 
    O(\Gamma_R \oplus \Gamma^*_R,q)/O(m) \times O(m).
\]
For $\delta = 0$ and $m=4$ this is exactly \pref{eq:MNfour}.

\subsection{$N=2$ SCFTs on tori}
To discuss mirror symmetry we need the more refined structure of an
$N=2$ SCFT. Recall that we will use the definition of an $N=2$
structure that involves a choice of the generators $J(z)$ and
$\bar{J}(\bar{z})$. So we will study the Teichmüller space
$\TMp_{N=2}$.  To define this space we slightly modify a definition
due to Kapustin and Orlov.
\begin{dfn}
\label{def:MNtwo}
The Teichmüller space of $N=2$ SCFTs on a torus of real dimension $m$ is
defined as
\[
  \TMp_{N=2} := \biggl \{ (\Gamma, I, J) \biggm | \negthickspace
    \begin{array}{l}
      \text{$\Gamma$ lattice of rank $m$, $I,J: \Gamma_\R \oplus
        \Gamma_\R^* \rightarrow \Gamma_\R \oplus \Gamma_\R^*$, 
        $I^2 = J^2 = -\Id$} \\
      \text{$[I,J] = 0$, $I^t q I = q$, $J^t q J = q$, $q I J$ pos.\ def.}
    \end{array} \negthickspace \biggr \}.
\]
\end{dfn}
As we will see below these data define two complex structures on the
torus $X$, which implies that the real dimension $m$ of the torus has
to be even for this space to be nonempty. Therefore we will often
write $2n$ instead of $m$, where $n$ is the complex dimension of the
torus.  Comparing this definition to \sref{def:MNone} one can easily
check that there exists a natural map $\pi: \TMp_{N=2} \rightarrow
\TM_{N=1}$ defined by $K=IJ$. This is the concrete realisation of the
map discussed in \pref{sec:gensit}.  The definition in~\cite{KO00}
contains one extra condition, so the corresponding Teichmüller space
is a subspace of the one defined here.  Anticipating a result below we
will denote this subspace by $\TMp_{N=2}^{\mathrm{geom}}$. It can be
defined in terms of $\TMp_{N=2}$ as follows
\begin{equation}
\label{eq:Mgeom}
  \TMp_{N=2}^{\mathrm{geom}} := \{ (\Gamma, I, J) \in \TMp_{N=2} \mid
  J(0 \oplus \Gamma_\R^*) \subset 0 \oplus \Gamma_\R^* \}.
\end{equation}
The extra condition allows us to find a nice geometrical description
of this Teichmüller space.
\begin{prop}
\label{prop:Mgeom}
  Geometrically the subspace $\TMp_{N=2}^{\mathrm{geom}}$ of
  $\TMp_{N=2}$ can be interpreted as follows
  \[
    \TMp_{N=2}^{\mathrm{geom}} = \biggl \{ (\Gamma, B, \omega, j)
    \biggm | \negthickspace
    \begin{array}{l}
      \text{$\Gamma$ lattice of rank $m$, $B, \omega: \Gamma_\R \rightarrow
      \Gamma_\R^*$ antisymmetric,} \\
      \text{$j: \Gamma_\R \rightarrow \Gamma_\R$, $j^2 = -\Id$, $j^t
      \omega j = \omega$, $\omega j$ pos.\ def.}
    \end{array} \negthickspace \biggr \}.
  \]
\end{prop}
\begin{proof}
  To prove the inclusion $\supset$ it suffices to check that $I$ and
  $J$ defined by
\begin{subequations}
\begin{align}
\label{eq:I}
  I &= \begin{twomatrix}
               \omega^{-1} B      &  -\omega^{-1} \\
         \omega + B \omega^{-1} B & -B \omega^{-1}
       \end{twomatrix}, \\
\label{eq:J}
  J &= \begin{twomatrix}
             j       &   0 \\
         B j + j^t B & -j^t
       \end{twomatrix}.
\end{align}
\end{subequations}
satisfy the conditions from \sref{def:MNtwo}. For the other inclusion,
note that the extra condition in~\pref{eq:Mgeom} ensures that we can
write $J=\smpmatrix{\alpha & 0 \\ \beta & \gamma}$. From $J^2 = -\Id$
it follows that $\alpha^2 = \gamma^2 = -\Id$. Using $J^t q J = q$, we
find that $\gamma = -\alpha^t$. Because of the map $\pi: \TMp_{N=2}
\rightarrow \TM_{N=1}$ we know that $K := IJ$ can be written in the
form \pref{eq:K}. So we can express $I = -KJ$ in terms of $\alpha$,
$\beta$, $G$ and $B$. From $I^2 = -\Id$ it follows that $\beta = B
\alpha + \alpha^t B$ (block in the upper right hand corner). If we now
define $\omega := - G \alpha$ and $j := \alpha$, then using again $I^2
= -\Id$ (block in the lower right hand corner) it follows that $j^t
\omega j = \omega$. This shows that $G = \omega j$, so $\omega j$ is
positive definite. Finally the antisymmetry of $\omega$ follows from
$I^t q I = q$.
\end{proof}
This description yields a geometrical interpretation because we can
regard $j$ as a complex structure on the torus compatible with the
metric $G$ in the sense that $j^t G j = G$. The antisymmetric matrix
$\omega$ is the corresponding Kähler form. Combining \sref{prop:MNone}
and~\ref{prop:Mgeom} we see that the restriction of $\pi$ to
$\TMp_{N=2}^{\mathrm{geom}}$ is given by $G = \omega j$. So the fibre
of $\pi|_{\TMp_{N=2}^{\mathrm{geom}}}$ over $(\Gamma,G,B)$ is just the
space of all complex structures $j$ satisfying $j^t G j = G$. Note
that $\pi(\TMp_{N=2}^{\mathrm{geom}})= \TM_{N=1}$, so the restriction
is still surjective.

Let us analyse the fibres of the map $\pi$ without imposing any extra
restrictions. One can check that the fibres are given by
\begin{equation}
\label{eq:fibres}
  \pi^{-1}(K) = \{ (-J K, J) \in \End(\Lambda_\R)^2 \mid \text{$J^2 =
  -\Id$, $[J,K] = 0$, $J^tqJ =q$} \}.
\end{equation}
Let $E_{\pm 1}$ be the eigenspaces of $K$ as discussed above. Because
$J$ and $K$ commute, the endomorphism $J$ should preserve these
eigenspaces. So $J$ can be defined by choosing complex structures on
$E_{\pm 1}$. Using the maps $R(G,B): \Lambda_\R \to
\Gamma_\R \oplus \Gamma_\R$ to write $J$ in block form we find
\begin{equation}
\label{eq:Jdiag}
  J = R(G,B)^{-1} \begin{twomatrix}
                     j_1 &  0 \\
                      0  & j_2
                  \end{twomatrix}
      R(G,B),
\end{equation}
where $j_{1,2} \in \End(\Gamma_\R)$ are complex structures on
$\Gamma_\R$.  Using \pref{eq:qdiag} we see that the last condition in
\pref{eq:fibres} boils down to $j_{1,2}$ being compatible with the
metric $G$. So if we define
\[
  S := \{ j \in \End(\Gamma_\R) \mid \text{$j^2 = -\Id_{\Gamma_\R}$,
  $j^t G j = G$} \},
\]
then the fibres of $\pi$ can be described as $\pi^{-1}(K) = S \times
S$. This leads to the following alternative description of the full $N=2$
Teichmüller space.
\begin{prop}
The full $N=2$ Teichmüller space has the following geometrical description
\label{prop:MNtwo}
\[
  \TMp_{N=2} = \biggl \{ (\Gamma,G,B,j_1,j_2) \biggm | \negthickspace
  \begin{array}{l}
    \text{$\Gamma$ lattice of rank $m$, $B,G: \Gamma_\R \to
      \Gamma_\R^*$, $B=-B^t$, $G=G^t$} \\
    \text{$G$ pos.\ def., $j_{1,2}: \Gamma_\R \rightarrow \Gamma_\R^*$, 
       $j_{1,2}^2 = -\Id$, $j_{1,2}^t G j_{1,2} = G$}
  \end{array} \biggr \}.
\]
In this notation the subset $\TMp_{N=2}^{\mathrm{geom}}$ is given by
the condition $j_1 = j_2$.
\end{prop}
\begin{proof}
  The alternative description of $\TMp_{N=2}$ follows from the
  discussion above, so we only need to prove the last statement.  If
  we write $J: \Gamma_\R \oplus \Gamma_\R^* \rightarrow \Gamma_\R
  \oplus \Gamma_\R^*$ in block form, the condition defining
  $\TMp_{N=2}^{\mathrm{geom}}$ corresponds to the vanishing of the
  block in the upper left which is a map from $\Gamma_\R^*$ to
  $\Gamma_\R$.  Using \pref{eq:Jdiag} one can easily check that this
  map is given by $\frac{1}{2}(j_1 - j_2)G^{-1}$. So we see that the
  geometrical case is precisely when $j_1 = j_2 = j$. In hindsight
  this justifies the use of the notation $\TMp_{N=2}^{\mathrm{geom}}$
  for the space defined in \pref{eq:Mgeom}.
\end{proof}
In other words fibrewise $\TMp_{N=2}^{\mathrm{geom}}$ is the
diagonal in $\TMp_{N=2}$.  Using $j^t G j = G$ one finds that
in that case $J$ is given by \pref{eq:J}. With $K$ fixed $I$ is
determined by $J$.  In the notation we established above we find
\begin{equation}
\label{eq:Idiag}
  I = R(G,B)^{-1} \begin{twomatrix}
                     -j_1 &  0 \\
                       0  & j_2
                  \end{twomatrix}
      R(G,B).
\end{equation}
Again one can check that in the geometrical case this reproduces
\pref{eq:I}. Altogether these results mean that we reproduce exactly
the diagram \pref{eq:Ntwodiag} that we found on general grounds in
\pref{sec:gensit}.

The underlying vector space of an $N=2$ vertex algebra is the same as
that of the corresponding $N=1$ vertex algebra. The $N=2$ structure is
defined by a set of fields $J(z)$, $G^\pm(z)$ and corresponding fields
in the right-moving sector which generate the $N=2$ superconformal
algebra (see \pref{eq:Ntwogens}). An $N=2$ isomorphism is an $N=1$
isomorphism that in addition preserves $J(z)$, $G^\pm(z)$ and their
right-moving counterparts.  This means that \emph{any} isomorphism $f:
V \rightarrow V'$ of $N=1$ vertex algebras that allow an $N=2$
structure can be lifted as an $N=2$ isomorphism by choosing an $N=2$
structure on $V$ and then using $f$ to transport it to $V'$.

As discussed above we can generalise the notion of $N=2$ isomorphisms
by including mirror morphisms. The most general possibility is
\begin{equation}
\label{eq:JepsLR}
\begin{split}
  f^{-1} J'(z) f &= \epsilon_L J(z),\\
  f^{-1} \bar{J}'(\bar{z}) f &= \epsilon_R \bar{J}(\bar{z}),
\end{split}
\end{equation}
where $(\epsilon_L, \epsilon_R) = (\pm 1, \pm 1)$. The map on
$G^\pm(z)$ and $\Gbar^\pm(\bar{z})$ has to satisfy the corresponding
conditions. For $(\epsilon_L, \epsilon_R) = (1,1)$ this reduces to the
definition of $N=2$ isomorphisms, whereas for $(\epsilon_L,
\epsilon_R) = (-1,1)$ and $(\epsilon_L, \epsilon_R) = (1,-1)$ these
conditions define left and right mirror morphisms (see
\sref{tab:genmor}).
\begin{table}
\caption{Generalised $N=2$ morphisms}
\label{tab:genmor}
\begin{center}
\begin{tabular}{lxxxxx}
\toprule
     & (\epsilon_L, &\epsilon_R) &   &I'      & &J'  & f^{-1}&J'(z)f &
     f^{-1}&\bar{J}'(\bar{z})f \\ 
\midrule
$N=2$ isomorphism     & (+1,&+1) &  g&Ig^{-1} &  g&Jg^{-1} &  &J(z) &
     &\bar{J}(\bar{z}) \\
left mirror morphism  & (-1,&+1) &  g&Jg^{-1} &  g&Ig^{-1} & -&J(z) &
     &\bar{J}(\bar{z})\\
right mirror morphism & (+1,&-1) & -g&Jg^{-1} & -g&Ig^{-1} &  &J(z) &
    -&\bar{J}(\bar{z})\\
complex conjugation   & (-1,&-1) & -g&Ig^{-1} & -g&Jg^{-1} & -&J(z) &
    -&\bar{J}(\bar{z})\\
\bottomrule
\end{tabular}
\end{center}
\end{table}

As above we can argue that any $N=1$ isomorphism $f: V \rightarrow V'$
can be lifted to a generalised $N=2$ morphism of type $(\epsilon_L,
\epsilon_R)$ by using \pref{eq:JepsLR} to define the $N=2$
structure on $V'$.  In the case of tori we can classify the
generalised $N=2$ morphisms using lattice isomorphisms.  The precise
relation is given by the following generalisation of a theorem
from~\cite{KO00}.
\begin{thm}
  Generalised $N=2$ morphisms of type $(\epsilon_L, \epsilon_R) = (\pm
  1, \pm 1)$ between the $N=2$ SCFTs $V_{\Gamma,I,J}$ and
  $V_{\Gamma',I',J'}$ correspond 1-1 to lattice isomorphisms $g:
  (\Gamma \oplus \Gamma^*,q) \rightarrow (\Gamma' \oplus
  {\Gamma'}^*,q')$ such that  $I'$ and $J'$ are given by \sref{tab:genmor}.
\end{thm}
Here we use $V_{\Gamma,I,J}$ to denote an $N=2$ SCFT. As a vector
space it is equal to $V_{\Gamma,K}$, but it comes with a choice of
$N=2$ generators. We will write $f_g^{(\epsilon_L, \epsilon_R)}$ for
the generalised $N=2$ morphism of type $(\epsilon_L, \epsilon_R)$
corresponding to the lattice isomorphism $g: (\Gamma \oplus
\Gamma^*,q) \rightarrow (\Gamma' \oplus {\Gamma'}^*,q')$. Kapustin and
Orlov formulated this classification only for the geometrical part of
the Teichmüller space and they only considered $N=2$ isomorphisms and
left mirror morphisms.  However, one can check that their proof
extends trivially to the more general case.  Like we did in the $N=1$
case we can define maps on the Teichmüller space for each lattice
isomorphism $g$ preserving the bilinear form.
\begin{dfn}
  Let $g: (\Gamma \oplus \Gamma^*,q) \rightarrow (\Gamma' \oplus
  {\Gamma'}^*,q')$ be a lattice isomorphism and let
  $(\epsilon_L,\epsilon_R) = (\pm 1, \pm 1)$, then we define the map
  $\mu_g^{(\epsilon_L, \epsilon_R)}: \TMp_{N=2}(\Gamma) \rightarrow
  \TMp_{N=2}(\Gamma')$ by $\mu_g^{(\epsilon_L, \epsilon_R)}(I,J) =
  (I',J')$, with $I'$ and $J'$ as defined in \sref{tab:genmor}.
\end{dfn}
Again we can choose a fixed lattice $\Gamma$ to study the duality
group and the moduli space. The duality group $\mathcal{G}$ is still
$O(\Gamma \oplus \Gamma^*,q)$. The action of an element $g \in
\mathcal{G}$ on $\TMp_{N=2}$ is given by $\mu_g^{(1,1)}$. An element
$(\epsilon_L, \epsilon_R, g)$ of the extended duality group
$\mathcal{G}' = O(\Gamma \oplus \Gamma^*,q) \times \Z_2 \times \Z_2$
acts as $\mu_g^{(\epsilon_L, \epsilon_R)}$. The simplest generalised
$N=2$ morphism are $f_{\Id}^{(\epsilon_L,\epsilon_R)}$. Together with
the $N=2$ isomorphisms they generate the extended duality group.

The generalised $N=2$ morphisms have the following functoriality
property.
\begin{prop}
\label{prop:genmorfunc}
  Let $g_i: (\Gamma_i \oplus \Gamma_i^*,q_i) \rightarrow (\Gamma_{i+1}
  \oplus \Gamma_{i+1}^*,q_{i+1})$ be lattice isomorphisms and let
  $(\epsilon_{L,i}, \epsilon_{R,i}) = (\pm 1, \pm 1)$, then we
  have the following functoriality property
\[
  f_{g_2}^{(\epsilon_{L,2},\epsilon_{R,2})} \circ 
  f_{g_1}^{(\epsilon_{L,1},\epsilon_{R,1})} = 
  f_{g_2 \circ g_1}^{(\epsilon_{L,2} \epsilon_{L,1}, \epsilon_{R,2} 
    \epsilon_{R,1})}.
\]
\end{prop}
This follows almost directly from \sref{prop:Nonemorfunc}, because a
generalised $N=2$ isomorphism is just an $N=1$ isomorphism satisfying
some extra conditions depending on the type $(\epsilon_L,
\epsilon_R)$. Using \pref{eq:JepsLR} it is easy to check that $f_{g_2
  \circ g_1}$ is of type $(\epsilon_{L,2} \epsilon_{L,1},
\epsilon_{R,2} \epsilon_{R,1}\epsilon_{R,1})$ if the $f_{g_i}$ are of
type $(\epsilon_{L,i}, \epsilon_{R,i})$. On this algebraic level this
is a rather trivial statement. It becomes more significant when we
consider the geometrical interpretation and the relation with
D-branes.

To find out how the complex structures $j_1$ and $j_2$ transform, one
can use the transformation rules for $I$ and $J$ from
\sref{tab:genmor}. Combining \pref{eq:Idiag} and \pref{eq:Jdiag} with
\pref{eq:gR}, one easily finds the formulae listed in
\pref{tab:trules}. Note that the signs for $j_1$ and $j_2$ coincide
with the ones for $J(z)$ and $\bar{J}(\bar{z})$. This is natural
because the fields $J(z)$ and $\bar{J}(\bar{z})$ are defined using the
complex structures $j_1$ and $j_2$ respectively.

\subsection{Geometrical interpretation}
\label{sec:geomint}
So far our description has been algebraic and based on a mathematical
description of SCFTs. In this section we try to interpret the
isomorphisms and generalised $N=2$ morphisms geometrically. As
discussed above they all correspond to a lattice isomorphisms $g:
(\Gamma \oplus \Gamma^*,q) \rightarrow (\Gamma' \oplus
{\Gamma'}^*,q')$. A general geometrical interpretation is complicated,
but we will discuss three classes of lattice isomorphisms that have a
nice geometrical interpretation.

\subsubsection*{Coordinate transformations}
The first class of lattice isomorphisms is $g=g_A = \smpmatrix{A & 0\\ 
  0 & A^{-t}}$, where $A: \Gamma \rightarrow \Gamma'$. These
correspond to isomorphisms of tori $\phi_A: X \rightarrow X': x
\mapsto A x$. So they have clear geometrical interpretation. In terms
of the metric and the B-field the map on Teichmüller space is given by
$\mu_g(G,B) = (A^{-t} G A^{-1}, A^{-t} B A^{-1})$, which is exactly
the expected transformation behaviour for a metric and a two-form.
  
Let us now consider the $N=2$ structure. Then for a given lattice
isomorphism $g$ there are four generalised $N=2$ morphisms
$f_g^{(\epsilon_L, \epsilon_R)}$ for $(\epsilon_L, \epsilon_R) = (\pm
1, \pm 1)$. However, if we want a geometrical interpretation, the map on
Teichmüller space should preserve the geometrical part of the Teichmüller
space. So if we start with a point $(B,G,j_1,j_2) \in \TMp_{N=2}$ with
$j_1=j_2$, then we should check using the transformation rules for
$j_1$ and $j_2$ from \sref{tab:trules} that this condition is
preserved. Because $\epsilon_L$ and $\epsilon_R$ determine the sign of
$j_1$ and $j_2$ respectively, the only thing that matters is the
relative sign. One can easily check that $\TMp_{N=2}$ is preserved
when $\epsilon_L \epsilon_R = 1$, i.e., for $N=2$ isomorphisms and for
complex conjugation (i.e., $(\epsilon_L,\epsilon_R) = (-1,-1)$).

\subsubsection*{Shifts in the B-field}
For the second class of lattice isomorphisms we assume that $\Gamma' =
\Gamma$. Then any antisymmetric map $C: \Gamma \rightarrow \Gamma^*$
defines a lattice isomorphism $g=g_C = \smpmatrix{\Id & 0\\ C & \Id}$.
In this case $\mu_g(G,B)= (G,B+C)$, so this lattice isomorphism
corresponds to a shift in the B-field. As above $f_g^{(\epsilon_L,
  \epsilon_R)}$ preserves the geometrical part of the Teichmüller
space if and only if $\epsilon_L \epsilon_R = 1$.

\subsubsection*{T-duality}
For the last class of lattice isomorphisms we consider lattices that
are the direct sum of two lattice $\Gamma = \Gamma_1 \oplus \Gamma_2$.
That allows us to define $\Gamma' = \Gamma_1 \oplus \Gamma_2^*$ and
the lattice isomorphism
\begin{equation}
\label{eq:gTdual}
  g_{\Gamma_1,\Gamma_2} =
       \begin{pmatrix}
          \Id_{\Gamma_1} &  0  &  0  &  0 \\
           0  &  0  &  0  & \Id_{\Gamma^*_2}\\     
           0  &  0  & \Id_{\Gamma^*_1} &  0 \\
           0  & \Id_{\Gamma_2} &  0  &  0
       \end{pmatrix},
\end{equation}
Geometrically this means that we write the torus as a product $X = X_1
\times X_2$, where $X_i = \Gamma_{i,\R}/\Gamma_i$. The second torus is
then given by $X' = X_1 \times X_2^*$, where $X_2^* =
\Gamma_{2,\R}^*/\Gamma_2^*$. This is T-duality in the subtorus $X_2 =
\Gamma_{2,\R}/\Gamma_2$.

For this class of lattice isomorphisms it is more complicated to check
if they preserve $\TMp_{N=2}^{\mathrm{geom}}$ as generalised $N=2$
morphisms. Let us write $x$ and $y$ for coordinates on $\Gamma_1$ and
$\Gamma_2$ and write $\check{x}$ and $\check{y}$ for the dual
coordinates. To analyse what happens to $I$ and $J$ in this case note
that $g$ is a permutation matrix interchanging $y$ and $\check{y}$. So
we can easily compute
\begin{equation}
\label{eq:Jp}
  g I g^{-1} = 
  \begin{pmatrix}
    I_{xx} & I_{x\check{y}} & I_{x\check{x}} & I_{xy} \\
    I_{\check{y}x} & I_{\check{y}\check{y}} 
      & I_{\check{y}\check{x}} & I_{\check{y}y} \\
    I_{\check{x}x} & I_{\check{x}\check{y}} 
      & I_{\check{x}\check{x}} & I_{\check{x}y} \\
    I_{yx} & I_{y\check{y}} & I_{y\check{x}} & I_{yy}
  \end{pmatrix}
\end{equation}
and similarly for $g J g^{-1}$. Here we use the variable names to
label the various components of linear maps.  In general it is
difficult to tell if a generalised $N=2$ morphism corresponding to
T-duality preserves $\TM_{N=1}^{\mathrm{geom}}$. However, two
important special cases can be analysed.

The first special case is when $\dim X_1 = \dim X_2$. In this case the
projection $\pi: X \rightarrow X_1$ defines a torus fibration of the
type needed for the Strominger-Yau-Zaslow conjecture describing mirror
symmetry (see~\cite{SYZ}). In their description mirror symmetry should
be T-duality in $X_2$. We can describe exactly when
$\mu_{g_{\Gamma_1,\Gamma_2}}^{(\epsilon_L,\epsilon_R)}$ with
$\epsilon_L \epsilon_R = -1$ preserves $\TMp_{N=2}^{\mathrm{geom}}$.
Note that the condition $\epsilon_L \epsilon_R = -1$ singles out the
left and right mirror morphisms.
\begin{prop}
\label{prop:SYZ}
Let $\Gamma = \Gamma_1 \oplus \Gamma_2$ and $\Gamma' = \Gamma_1 \oplus
\Gamma_2^*$ with $\rk \Gamma_1 = \rk \Gamma_2$, then
$\mu_{g_{\Gamma_1,\Gamma_2}}^{(\epsilon_L, \epsilon_R)}$ with
$\epsilon_L \epsilon_R = -1$ maps $(B,\omega,j) \in
\TMp_{N=2}^{\mathrm{geom}}(\Gamma)$ to
$\TMp_{N=2}^{\mathrm{geom}}(\Gamma')$ if and only if $\omega|_{X_2} =
0 = B|_{X_2}$.
\end{prop}
\begin{proof}
  Let $(I,J)$ correspond to $(B,\omega,j) \in
  \TMp_{N=2}^{\mathrm{geom}}(\Gamma)$ according to \pref{eq:I} and
  \pref{eq:J}. Using $\epsilon_L \epsilon_R = -1$ and the expressions
  for $I'$ and $J'$ from \sref{tab:genmor}, we find $(I',J') =
  \mu_{g_{\Gamma_1,\Gamma_2}}^{(\epsilon_L, \epsilon_R)}(I,J) =
  \epsilon_R (g J g^{-1}, g I g^{-1})$. The condition that $(I',J')$
  be in $\TMp_{N=2}^{\mathrm{geom}}(\Gamma')$ is equivalent to $J'$
  satisfying $J'(0 \oplus {\Gamma'}^*) \subset 0 \oplus {\Gamma'}^*$.
  Because $\epsilon_R J'$ is given by \pref{eq:Jp}, this condition is
  equivalent to
\[
  \begin{twomatrix}
    I_{x\check{x}}     &    I_{xy} \\
    I_{\check{y}\check{x}} & I_{\check{y}y}
  \end{twomatrix} = 0.
\]
From $I_{x\check{x}} = 0$ it follows using \pref{eq:I} that
$\omega_{\check{y}y} = 0$. Here it is essential that $\Gamma_1$ and
$\Gamma_2$ have equal rank.  Combining this with $I_{xy} = 0$ we find
$B_{\check{y}y} = 0$. It can easily be checked that these two
conditions also guarantee that the remaining components vanish.
\end{proof}
The condition $\omega|_{X_2} = 0$ can be interpreted geometrically as
the requirement that the fibres of the SYZ-fibration be Lagrangian.
The corresponding condition for the B-field does not have such a nice
interpretation. This proposition also gives a nice explanation why the
cases when these conditions are not fulfilled are so difficult to
understand.  Mirror symmetry then takes us to a non geometrical part
of the Teichmüller space where many of our standard assumptions will
cease to be valid.  This is probably why so far all attempts to
describe mirror symmetry on higher dimensional tori contain some
assumptions that exclude those cases.

A remark is in order about what we call mirror symmetry. Above we
associated left and right mirror morphisms to any lattice isomorphism
preserving the bilinear form. This seems to be at odds with the
SYZ-conjecture, which states that mirror symmetry corresponds to
T-duality in the fibres of a very special torus fibration.  A full
reconciliation of these two points of view will probably require a
careful analysis of the arguments in~\cite{SYZ} leading to the
SYZ-conjecture. The key point seems to be that mirror morphisms
generally do not preserve $\TMp_{N=2}^{\mathrm{geom}}$. The above
proposition shows that for the class of mirror morphisms corresponding
to the SYZ-conjecture there is at least a big subspace of
$\TMp_{N=2}^{\mathrm{geom}}(\Gamma)$ that is mapped to
$\TMp_{N=2}^{\mathrm{geom}}(\Gamma')$. The argument in~\cite{SYZ}
depends on the standard geometrical interpretation of D-branes which
is only possible in the geometrical part of the Teichmüller space.
That might restrict the class of mirror morphism to which the argument
applies to fibrewise T-dualities as predicted by the SYZ-conjecture.

It is possible to define new Teichmüller spaces that are invariant under a
fixed $g=g_{\Gamma_1,\Gamma_2}$.
\begin{align*}
  \TMp_{N=2}^{\mathrm{SYZ}} &:= \{ (B,\omega,j) \in 
    \mathcal{M}_{N=2}^{\mathrm{geom}} \mid
    B_{\check{y}y} = \omega_{\check{y}y} = 0 \} \\
  \TM_{N=1}^{\mathrm{SYZ}} &:= \{ (B,G) \in
    \mathcal{M}_{N=1}^{\mathrm{geom}} \mid B_{\check{y}y}  = 0 \}
\end{align*}
Note that $\TMp_{N=2}^{\mathrm{SYZ}}$ depends on the choice of the
SYZ-fibration.  All these spaces can be put together in the following
diagram
\[
\xymatrix{
  \TMp_{N=2}^{\mathrm{SYZ}} \ulab{4n^2} 
    \ar@{^{(}->}[r] \ar[d] &
  \TMp_{N=2}^{\mathrm{geom}} \ulab{4n^2+n(n-1)}
  \ar@{^{(}->}[r] \ar[dr] &
  \TMp_{N=2} \ulab{4n^2+2n(n-1)} \ar[d] &
  S \times S \ar[l] \ulab{2 \times n(n-1)} \\
  \TM_{N=1}^{\mathrm{SYZ}} \dlab{4n^2-\tfrac{1}{2}n(n-1)}
    \ar@{^{(}->}[rr] & &
  \TM_{N=1} \dlab{4n^2}}
\]
This diagram is an extension of the one in \pref{eq:Ntwodiag}. Recall
that the Betti numbers of a torus $X$ of complex dimension $n$ are
given by $h^{p,q}(X) = \binom{n}{p} \binom{n}{q}$.  Then it is easy to
check that the dimensions agree with the general formulae from
\pref{eq:Ntwodiag}.  One can also check that this diagram commutes. It
is tempting to restrict to $\TMp_{N=2}^{\mathrm{SYZ}}$ and
$\TM_{N=1}^{\mathrm{SYZ}}$, because the geometrical interpretation is
clearer. However, to obtain a completely general description it is
necessary to work with the most general Teichmüller spaces.

The second case that can be analysed is when $\Gamma_2 = \Gamma$. This
corresponds to a Fourier-Mukai transform on the whole torus. In this
case we want to see when $f_g^{(\epsilon_L, \epsilon_R)}$ with
$\epsilon_L \epsilon_R = 1$ preserve $\TMp_{N=2}^{\mathrm{geom}}$. So
suppose that $j_1 = j_2 = j$, then the condition $j_1' = j_2'$ is
equivalent to
\[
  E j E^{-1} = E^t j E^{-t}.
\]
Note that $G$ is Kähler so $G j = -j^t G$. If we suppose that $B$ is a
$(1,1)$-form, then we also have $B j = -j^t B$, so $E j = -j^t E$ and
$E^t j = -j^t E^t$. This suffices to show that the above condition for
$j_1' = j_2'$ is fulfilled. It is not clear to me if this condition is
also necessary.

On this level one can easily check that for a fixed decomposition
$\Gamma = \Gamma_1 \oplus \Gamma_2$, doing a fibrewise T-duality in
the $\Gamma_1$-directions and then in the $\Gamma_2$-directions (or
the other way around) yields full Fourier-Mukai transform. If one
starts with a SYZ-decomposition and lifts both fibrewise T-dualities
as mirror morphisms, then this is exactly the situation of the
conjecture discussed in Section~4.5.4 of~\cite{Enc00}.

\section{D-branes}
\subsection{Gluing matrices}
In~\cite{OOY96} the authors analyse in detail the geometrical
interpretation of D-branes. In the terminology developed above their
analysis is restricted to $\TMp_{N=2}^{\mathrm{geom}}$.  Let us try to
extend their description of D-branes to one that is valid over the
entire Teichmüller space. Of course, it is likely to be difficult to
find a completely geometrical interpretation outside
$\TMp_{N=2}^{\mathrm{geom}}$. The authors of~\cite{OOY96} used a gluing
field $R$ to describe boundary conditions (see
also~\cite{Sch02,RS98,RS97}). We will use the same description. In
general $R$ is a field, but for a torus we can use a constant matrix,
which we will call the \emph{gluing matrix}. As we will discuss below,
this amounts to restricting to affine D-brane.  On the boundary of the
worldsheet one imposes the following conditions
\begin{subequations}
\label{eq:R}
\begin{align}
\label{eq:RX}
\partial X(z)|_{\sigma=0,\pi} 
  &= R \bar{\partial}X(\bar{z})|_{\sigma=0,\pi}, \\
\label{eq:Rpsi}
\psi(z)|_{\sigma=0,\pi} &= \eta R \bar{\psi}(\bar{z})|_{\sigma=0,\pi}.
\end{align}
\end{subequations}
Here we introduced a parameter $\eta$, which can have the values $\pm
1$. Because GSO-invariant states are linear combinations of states for
both values of $\eta$ (see pages 15--16 in~\cite{Gab02}), this
parameter is necessary in the general theory. However, in this paper
it does not play an important role and we will just carry it along.
In the expression above we used the usual complex coordinate $z =
\exp(\tau+i\sigma)$ of $z$ in terms of $\tau$ and $\sigma$. This means
that the strip $\{ (\tau, \sigma) \mid 0 < \sigma < \pi \}$ is
identified with the upper half plane.  In~\cite{ALZ01,ALZ02} more
general boundary conditions are discussed containing extra terms in
the equations \pref{eq:RX} for the bosonic fields.  However, in our
case their boundary conditions reduce to \pref{eq:R}, because $R$ is
constant and we use flat coordinates on the torus.

Using the formulae from \sref{tab:trules} one can easily check that
after an isomorphism of $N=1$ SCFTs corresponding to a lattice
isomorphism $g = \smpmatrix{a & b \\ c & d}$ the boundary conditions
for the new fields are given by \pref{eq:R} with $R$ replaced by
\[ 
  R' = (a + b E) R (a - bE^t)^{-1}.
\]
In~\cite{OOY96} the authors give a different formula for the special
case where $g$ corresponds to a T-duality transformation. They
identify the lattices $\Gamma$, $\Gamma^*$, $\Gamma'$, and
${\Gamma'}^*$ and assume implicitly that the background field $E$ is
simply $\Id$. They claim that a T-duality transformation is then given
by a symmetric matrix $T$ satisfying $T^2 = \Id$. If $H \subset
\Gamma_\R$ is the subspace that is dualised, then we can define $T$ by
the requirements $T|_H = -\Id_H$ and $T|_{H^\perp} = \Id_{H^\perp}$.
The claim of~\cite{OOY96} is that in terms of $T$ the transformation
of the gluing matrix is given by $R' = RT$. To compare to our
description, we have to find the lattice isomorphism $g$ corresponding
to this T-duality transformation. Generalising \pref{eq:gTdual}, we
find $g = \smpmatrix{a & b \\ b & a}$, where $a = \frac{1}{2}(\Id+T)$
and $b = \frac{1}{2}(\Id-T)$.  With $E = \Id$ it then follows easily
that both formulae agree.

The matrix $R$ should satisfy several conditions in order to define a
boundary condition for an $N=1$ SCFT\@. In~\cite{ALZ02} three
conditions are listed, which we will discuss in turn. The mathematical
formulation of the first condition is rather natural.
\begin{cond}[orthogonality]
\label{cond:ortho}
  The matrix $R$ should be orthogonal with respect to the metric $G$,
  i.e., $R^t G R = G$.
\end{cond}
Physically this condition corresponds to the requirement $L(z) =
\bar{L}(\bar{z})$ on the boundary. As was argued in~\cite{OOY96} the
$-1$ eigenspace of $R$ should correspond to directions orthogonal to
the D-brane. So let $V \subset \Gamma_\R$ be the $-1$ eigenspace of
$R$ and let $W$ be its orthogonal complement with respect to the
metric $G$.  Using $V$ and $W$ we can define $V^* := G(V)$ and $W^* :=
G(W)$. Note that $\Gamma^*_\R = V^* \oplus W^*$. Equivalently, we can
use the canonical pairing $\langle \cdot, \cdot \rangle: \Gamma_\R
\otimes \Gamma^*_\R \rightarrow \R$ and define $V^* = W^\perp := \{ u
\in \Gamma^* \mid \forall w \in W: \langle w, u \rangle =0 \}$ and $W^*
= V^\perp$.

With respect to the direct sum decomposition $\Gamma_\R = V \oplus W$,
the matrix $R$ has the block diagonal form
\begin{equation}
\label{eq:stdR}
  R = \begin{twomatrix}
        -\Id &   0 \\
          0  & \tilde{R}
      \end{twomatrix}.
\end{equation}
By definition the map $\tilde{R}: W \rightarrow W$ does not have $-1$ as
an eigenvalue.  It follows that there exists a unique map
$\mathcal{F}: W \rightarrow W^*$ such that $\tilde{R} = (\tilde{G} -
\mathcal{F})^{-1} (\tilde{G} + \mathcal{F})$, where $\tilde{G}: W
\rightarrow W^*$ is the restriction of $G$ to $W$. In fact one can easily
compute that $\mathcal{F}$ is given by $\mathcal{F} =
\tilde{G}(\tilde{R} - \Id)(\tilde{R} + \Id)^{-1}$. We also know that
$\tilde{R}$ is orthogonal with respect to $\tilde{G}$, which can be
written as $\tilde{G}^{-1} = \tilde{R}^{-1} \tilde{G}^{-1}
\tilde{R}^{-t}$. Using this one can easily show that $\mathcal{F}$ is
antisymmetric. Here we use the decompositions $\Gamma_\R = V \oplus W$
and $\Gamma^*_\R = V^* \oplus W^*$ to identify the transpose of a map
$W \rightarrow W^*$ with a map $W \rightarrow W^*$ again.

Geometrically $W$ can be regarded as a subspace of the tangent space
to $X = \Gamma_\R/\Gamma$. The submanifold on which the D-brane lives
should have $W$ as its tangent space. In general we need an
integrability condition to ensure that there exists at least locally a
submanifold $S \subset X$ such that its tangent space is $W$. This is
the second condition from~\cite{ALZ02}. In the present case $W$ is
constant and the integrability condition is automatically fulfilled.
However, this condition only guarantees the local existence of the
submanifold $S$. For the torus we can do better by imposing the
following condition which ensures the global existence of $S$ as a
submanifold of $X$.
\begin{cond}[rationality]
\label{cond:rat1}
  Let $W$ be the orthogonal complement with respect to $G$ of the
  $-1$ eigenspace of $R$, then $W \cap \Gamma$ should have rank equal
  to the real dimension of $W$.
\end{cond}
If this condition is not fulfilled, the integral manifold will not be
compact. More precisely, it will be dense in a submanifold of $X$ with
dimension higher than the dimension of $W$. For the 2-dimensional
torus $\R^2/\Z^2$ this is familiar and corresponds to lines with
irrational slope which are dense in $\R^2/\Z^2$. Note that because
$V^* = W^\perp$, this condition is equivalent to the condition that
$V^* \cap \Gamma^*$ have rank equal to the real dimension of $V^*$.

The third condition allows us to interpret $\tilde{R}$ as defined
above in more detail.  This requires a slight modification of the
condition as formulated in~\cite{ALZ02} to allow for D-branes with a
nontrivial field strength. The condition from~\cite{ALZ02} can be
written as $\tilde{R} = \tilde{E}^{-1} \tilde{E}^t = (\tilde{G} +
\tilde{B})^{-1} (\tilde{G} - \tilde{B})$. Here $\tilde{B}: W
\rightarrow W^*$ is obtained from $B$ by restricting $B$ to $W$ and
then projecting to $W^*$. It can be checked using $V^* = W^\perp$ that
$\langle w, \tilde{B} w' \rangle = \langle w, B w' \rangle$ for all
$w, w' \in W$. In this sense $\tilde{B}$ does define the restriction
of $B$ to $W$ as a bilinear form. 

When the field strength $F$ of the connection on the D-brane is
nonzero, the condition discussed above has to be modified. To do so we
define $\mathcal{F} := F - \tilde{B}$ and replace $\tilde{B}$ by
$-\mathcal{F}$. For line bundles on a torus the field strength has to
be integral as a bilinear form on the lattice defining the torus.  As
will be discussed below this condition can be relaxed to the
requirement that $F$ be rational. This leads to the second rationality
condition.
\begin{cond}[rationality]
\label{cond:rat2}
If we write $R$ as in \pref{eq:stdR}, then there exists an
antisymmetric matrix $\mathcal{F}$ such that $\tilde{R} = (\tilde{G} -
\mathcal{F})^{-1} (\tilde{G} + \mathcal{F})$. As an antisymmetric
bilinear form $F := \mathcal{F} + \tilde{B}$ should map $W \cap \Gamma
\otimes W \cap \Gamma$ to $\Q$.
\end{cond}
The condition in~\cite{ALZ02} amounts to requirement that the field
strength $F$ vanish, so our condition certainly is a generalisation of
that condition. 

\begin{table}
\caption{Transformation rules}
\label{tab:trules}
\begin{center}
\begin{tabular}{lxx}
\toprule
&\multicolumn{2}{c}{Background data} & \multicolumn{2}{c}{Fields} \\
\midrule
$N=1$&     E' &= (c+dE)(a+bE)^{-1}
&       f_g^{-1} \partial X' f_g &= (a+bE) \partial X \\
&-{E'}^t &= (c-dE^t)(a-bE^t)^{-1}
& f_g^{-1} \bar{\partial} X' f_g &= (a-bE^t) \bar{\partial} X \\
&     G' &= (a+bE)^{-t} G (a+bE)^{-1}
&             f_g^{-1} \psi' f_g &= (a+bE) \psi \\
&        &= (a-bE^t)^{-t} G (a-bE^t)^{-1}
&       f_g^{-1} \bar{\psi}' f_g &= (a-bE^t) \bar{\psi} \\
\midrule
$N=2$
&   j_1' &= \epsilon_L (a+bE) j_1 (a+bE)^{-1} \\
&   j_2' &= \epsilon_R (a-bE^t) j_2 (a-bE^t)^{-1} \\
& \omega_1' &= \epsilon_L (a+bE)^{-t} \omega_1 (a+bE)^{-1} \\
& \omega_2' &= \epsilon_R (a-bE^t)^{-t} \omega_2 (a-bE^t)^{-1} \\
\midrule
D-branes & R' &= (a+bE)R(a-bE^t)^{-1} \\
\bottomrule
\end{tabular}
\end{center}
\end{table}

The logical next step would be to check that these conditions are well
behaved under $N=1$ isomorphisms. The orthogonality condition is easy
to check. From the transformation rules in \sref{tab:trules} it is
immediate that ${R'}^t G' R' = G'$. The rationality conditions are
more complicated, because the decomposition $\Gamma_\R = V \oplus W$
changes under an isomorphism of $N=1$ SCFTs.  It can be shown in
examples that the two rationality conditions mix in the sense that one
of the rationality conditions after an isomorphism may depend on both
rationality conditions before the isomorphism. Therefore we will
describe in the next section an alternative description of D-branes
which replaces the two separate rationality conditions with a single
rationality condition. This will enable us to show that isomorphisms
also preserve the rationality conditions.

\subsection{Boundary states}
An important way to describe D-branes is using boundary states. These
are states in an extension of the closed CFT Hilbert space such that
expectation values of bulk operators $\phi_1$, \dots, $\phi_k$ are
given by
\[
  \langle \phi_1 \dots \phi_k \rangle_\alpha 
  = \langle \phi_1 \dots \phi_k \lVert \alpha \rAngle,
\]
where we use $\alpha$ to label the D-brane.  These states are linear
combinations of the so-called Ishibashi states. For the torus we can
denote the Ishibashi states as $\lvert R,w,m \rAngle$ using the gluing
matrix $R$ and $(w,m) \in \Gamma \oplus \Gamma^*$ to label them. The
defining condition for the Ishibashi states is
\begin{equation}
\label{eq:bdyglue}
  (\alpha_\ell + R \bar{\alpha}_{-\ell}) \lvert R,w,m \rAngle = 0.
\end{equation}
This equation corresponds to \pref{eq:RX} after translation from the
upper half plane to the complement of the unit disk
(see~\cite{Sch02}).  For a torus we can write down explicit solutions
of these equations
\begin{equation}
\label{eq:Ishi}
  \lvert R,w,m \rAngle = \exp \bigl (-\sum_{\ell=1}^\infty
  \frac{1}{\ell} G(\alpha_{-\ell}, R \bar{\alpha}_{-\ell} ) \bigr )
  \lvert w,m \rangle.
\end{equation}
These solutions are straightforward generalisations of the solutions
for the 1-dimensional case (see e.g.,~\cite{Gab02,Sch02}). Using the
commutation relations for the $\alpha^\mu_\ell$ and the
$\bar{\alpha}^\mu_\ell$ we can check that these states satisfy
\pref{eq:bdyglue} for $\ell \ne 0$. To check that condition for $\ell
= 0$, recall that $V_{w,m}$ are eigenspaces of $\alpha_0$ and
$\bar{\alpha}_0$. The eigenvalues of $\alpha_0$ and $\bar{\alpha}_0$
are $G^{-1} k_L$ and $G^{-1} k_R$ respectively, where $k_L$ and $k_R$
have been defined in terms of $(w,m) \in \Gamma \oplus \Gamma^*$ in
\pref{eq:kLR}. Substituting this into the requirement
\pref{eq:bdyglue} for $\ell = 0$ we find
\begin{equation}
\label{eq:Ldef}
  G^{-1}(Gw - Bw + m)= -R G^{-1}(-Gw - Bw + m).
\end{equation}
This condition defines a lattice $L_{R,K} \subset \Gamma \oplus \Gamma^*$
depending on the gluing matrix $R$ and the background fields $G$ and
$B$. For a fixed background this lattice turns out to be equivalent to
the gluing matrix $R$ as we see in the following theorem.
\begin{thm}
\label{thm:LR}
  Let $G$ and $B$ be fixed, then a gluing matrix $R$ satisfying the
  conditions~\ref{cond:ortho}, \ref{cond:rat1}, and~\ref{cond:rat2} is
  equivalent to a sublattice $L \subset \Gamma \oplus \Gamma^*$
  satisfying the conditions
\begin{gather*}
 q|_L = 0,\\
\rk L = \rk \Gamma.
\end{gather*}
\end{thm}
\begin{proof}
  Given a gluing matrix $R$ satisfying the
  conditions~\ref{cond:ortho}, \ref{cond:rat1}, and~\ref{cond:rat2},
  we can take $L$ to be the sublattice $L_{R,K} \subset \Gamma \oplus
  \Gamma^*$ defined by \pref{eq:Ldef}. Using $\Gamma_\R = V \oplus W$
  we can rewrite this equation as two equations, one for the
  $V$-component and one for the $W$-component. These two equations can
  be simplified further using \pref{eq:stdR} and $\tilde{R} = (\tilde{G}
  - \mathcal{F})^{-1}(\tilde{G} + \mathcal{F})$. After some rewriting
  we find
  \begin{align}
  \label{eq:LRV}
    w_V &= 0, \\
    (m - B w)_{W^*} &= \mathcal{F} w_W. \notag
  \end{align}
  Here the subscripts $V$, $W$, $V^*$, and $W^*$ label the components
  and we also used $\Gamma^*_\R = V^* \oplus W^*$.  The first equation
  implies that $w \in W$, so we can drop that equation and use the
  second equation to describe $L$ as a sublattice of $(W \cap \Gamma)
  \oplus \Gamma^*$. The fact that $w \in W$ allows us to omit the
  subscript $W$ and to replace $(B w)_{W^*}$ by $\tilde{B} w$. If we
  also use $\mathcal{F} = F - \tilde{B}$, we are left with the
  following equation defining $L$ as a sublattice of $(W \cap \Gamma)
  \oplus \Gamma^*$
  \begin{equation}
  \label{eq:LRW}
    m_{W^*} = F w.
  \end{equation}
  This vector equation stands for $k$ independent equations. More
  explicitly, we can choose an integral basis $\{ e_i \}$ for $W \cap
  \Gamma$. Then for every $i=1,\dots,k$ we obtain an equation
  \[
    \langle e_i, m_{W^*} \rangle = \langle e_i, m \rangle
    = \langle e_i, F w \rangle.
  \]
  Because of the second rationality condition~\ref{cond:rat2}, these
  equations are linear with rational coefficients. So the lattice $L$
  is a codimension $k$ sublattice in the $(k+\rk \Gamma)$-dimensional
  lattice $(W \cap \Gamma) \oplus \Gamma^*$. Therefore the rank of the
  lattice $L_{R,K}$ is indeed equal to the rank of $\Gamma$.
  
  To see that the restriction of $q$ to $L$ vanishes we compute
  \[
  \begin{split}
    q \bigl (\smpmatrix{w^{\vphantom{\prime}}\\m^{\vphantom{\prime}}}, 
             \smpmatrix{w'\\m'} \bigr ) 
    &= \langle w, m' \rangle + \langle w', m \rangle 
     = \langle w_V^{\vphantom{\prime}}, m_{V^*}' \rangle 
       + \langle w_W^{\vphantom{\prime}}, m_{W^*}' \rangle 
       + \langle w_V', m_{V^*}^{\vphantom{\prime}} \rangle
       + \langle w_W', m_{W^*}^{\vphantom{\prime}} \rangle\\
    &= \langle w_W^{\vphantom{\prime}}, F w_W' \rangle 
       + \langle w_W', F w_W^{\vphantom{\prime}} \rangle = 0.
  \end{split}
  \]
  Here we used the antisymmetry of $F$ in the final step. This shows
  that we can associate a sublattice $L$ satisfying the conditions from
  the statement of the theorem to any gluing matrix $R$ satisfying the
  conditions~\ref{cond:ortho}, \ref{cond:rat1}, and~\ref{cond:rat2}.
  
  Conversely, if we start with such a sublattice $L$, then we find a
  sublattice of $\Gamma$ by projecting $L \subset \Gamma \oplus
  \Gamma^*$ to $\Gamma$. Tensoring this sublattice with $\R$ yields a
  subspace $W$ satisfying $\rk (W \cap \Gamma) = \dim_\R W$.  Using
  $G$ we can define $V$ as the orthogonal complement of $W$, and $V^*$
  and $W^*$ as $G(V)$ and $G(W)$ respectively. As above, we can express
  $q$ in terms of the components. Let $(w,m)$ and $(w',m')$ be
  elements of the lattice $L$. Because $w_V^{\vphantom{\prime}}$ and
  $w_V'$ vanish, we find
  \[
    q \bigl (\smpmatrix{w^{\vphantom{\prime}}\\m^{\vphantom{\prime}}}, 
             \smpmatrix{w'\\m'} \bigr ) 
    = \langle w_W^{\vphantom{\prime}}, m_{W^*}' \rangle 
      + \langle w_W', m_{W^*}^{\vphantom{\prime}} \rangle.
  \]
  As $q|_L$ vanishes, this implies $\langle w_W',
  m_{W^*}^{\vphantom{\prime}} \rangle = -\langle
  w_W^{\vphantom{\prime}}, m_{W^*}' \rangle$. We can choose a basis
  $e_i \in W \cap \Gamma$ of $W$, such that there exist $f_i \in
  \Gamma^*$ satisfying $(e_i,f_i) \in L$. Substituting $(e_i,f_i)$ for
  $(w',m')$, we find $\langle e_i, m_{W^*}^{\vphantom{\prime}} \rangle
  = -\langle w_W^{\vphantom{\prime}}, f_{i,W^*} \rangle$. It follows
  that $m_{W^*}$ is determined by $w_W$. Because $L$ is a sublattice,
  we must have $m_{W^*} = F w_W$ for some linear map $F: W \rightarrow
  W^*$. The map $F$ has to be antisymmetric for $q|_L$ to vanish. The
  fact that $L$ is a lattice of rank $\rk \Gamma$ ensures that $F$ is
  rational.  Using $F$ we can define $\mathcal{F}$ as $F-\tilde{B}$
  and $\tilde{R}$ as $(\tilde{G} - \mathcal{F})^{-1}(\tilde{G} +
  \mathcal{F})$. Finally, the gluing matrix $R$ can be defined as the
  block matrix \pref{eq:stdR} with respect to $\Gamma_\R = V \oplus
  W$.
\end{proof}
So instead of using a gluing matrix $R$ to describe a D-brane, we
might as well use the corresponding lattice $L = L_{R,K}$. The
rationality conditions on $R$ correspond to $\rk L = \rk \Gamma$. Note
that this is the maximal rank for a lattice $L$ satisfying $q|_L = 0$.
This is a much more manageable condition when checking the behaviour
under $N=1$ isomorphisms. However, before we can do that, we have to
determine how the lattice $L$ transforms under $N=1$ isomorphisms.
\begin{prop}
\label{prop:transLR}
  Let $L_{R,K}$ be the sublattice from \sref{thm:LR} corresponding to a
  gluing matrix $R$ and let $g: (\Gamma \oplus \Gamma^*,q) \rightarrow
  (\Gamma' \oplus {\Gamma'}^*,q')$ be a lattice isomorphism, then the
  sublattice $L_{R',K'}$ corresponding to transformed gluing matrix $R'$
  is given by
  \[
    L_{R',K'} = g(L_{R,K}).
  \]
\end{prop}
\begin{proof}
  The condition \pref{eq:Ldef} defining $L_{R,K}$ can be written as
  \[
    \begin{twomatrix} G^{-1} & 0 \\ 0 & -R G^{-1} \end{twomatrix}
    \begin{twomatrix} E^t & \Id \\ -E & \Id \end{twomatrix}
    \begin{twomatrix} w \\ m \end{twomatrix}
    = \begin{twomatrix} \Id & 0 \\ 0 & R \end{twomatrix}
      \mathcal{R}(G,B) \begin{twomatrix} w \\ m \end{twomatrix}
    \in \Delta,
  \]
  where $\Delta$ is the diagonal in $\Gamma_\R \oplus
  \Gamma_\R^*$. In other words the lattice $L_{R,K}$ can be defined as
  \[
    L_{R,K} = \mathcal{R}(G,B)^{-1} \begin{twomatrix} \Id & 0 \\ 
          0 & R^{-1} \end{twomatrix} \Delta.
  \]
  Applying the lattice isomorphism $g$ to both sides, we obtain
  \[
  \begin{split}
    g(L_{R,K}) &= g \mathcal{R}(G,B)^{-1} \begin{twomatrix} \Id & 0 \\ 
          0 & R^{-1} \end{twomatrix} \Delta
     = \mathcal{R}(G',B')^{-1} \begin{twomatrix} (a+bE) & 0 \\ 
          0 & (a-bE^t) R^{-1} \end{twomatrix} \Delta \\
    &= \mathcal{R}(G',B')^{-1} \begin{twomatrix} \Id & 0 \\ 
          0 & {R'}^{-1} \end{twomatrix} \Delta
     = L_{R',K'}.
  \end{split}
  \]
  The second equality is based on \pref{eq:gR}. For the next step we
  used the definition of $R'$ and the fact that $\diag(a+bE,
  a+bE)$ leaves $\Delta$ invariant.
\end{proof}
It is clear that this transformation of $L$ preserves the conditions
from \sref{thm:LR} defining the subspace $L$. It follows that $N=1$
isomorphisms must also preserve the conditions defining a gluing
matrix $R$. Note that although we motivated the introduction of the
sublattice $L$ by studying Ishibashi states, this theorem is
independent of that and is mathematically rigorous.

The Ishibashi state defined above is for a purely bosonic theory. In
a supersymmetric theory there are additional requirements related to
the fermions
\begin{equation}
\label{eq:bdygluef}
  (\psi_r + i \eta R \bar{\psi}_{-r}) \lvert R, w, m, \eta \rAngle.
\end{equation}
Here we use again the parameter $\eta = \pm 1$ introduced above. As in
the bosonic case, this equation can be obtained from the corresponding
boundary condition \pref{eq:Rpsi} on the upper half plane.  Also in
this case we can write down explicit solutions
\[
  \lvert R, \eta, w, m \rAngle = \exp \bigl (-\sum_{\ell=1}^\infty
  \frac{1}{\ell} G(\alpha_{-\ell}, R \bar{\alpha}_{-\ell} ) 
  -i\eta \sum_{\ell=0}^\infty G(\psi_{-\ell-\frac{1}{2}}, 
   R \bar{\psi}_{-\ell-\frac{1}{2}} )\bigr ) \lvert w, m \rangle.
\]
However, in the sequel we will continue to use the bosonic
expressions, because they already contain all the information we need.
To do the supersymmetric case properly we would have to go into things
like the GSO-projection, which would lead us too far away
(see~\cite{Gab00,Gab02}).

Every boundary state has to satisfy the gluing conditions
\pref{eq:bdyglue} (and \pref{eq:bdygluef} in the fermionic case) and
can therefore be written as an infinite linear combination of
Ishibashi states. There are additional conditions that boundary states
have to satisfy, namely the Cardy condition and the sewing conditions.
However, it is unclear if all these conditions together are
sufficient. Therefore, we will refrain from studying these extra
conditions in detail and instead write down explicit expressions for
boundary states as linear combinations of Ishibashi states. It seems
quite plausible that these expressions actually define boundary
states, but the evidence we have is quite scarce. Firstly, the
expressions are straightforward generalisations of well known
expressions from the literature. Secondly, they lead to a description of
D-branes that agrees quite well with the expectations based on the
geometrical description of D-branes.

Let us now, as promised, return to the bosonic theory and write down
an explicit expression for a boundary state as a linear combination of
Ishibashi states
\begin{equation}
\label{eq:bdystate}
  \lVert R, \xi \rAngle = \sum_{(w,m) \in L} e^{-2 \pi i q((w,m), \xi)}
  \lvert R, w,m \rAngle.
\end{equation}
Here we introduce some extra data in the form of a vector $\xi \in
\Gamma_\R \oplus \Gamma_\R^*$. First note that because $q|_L$
vanishes, we can regard $\xi$ as an element of $\Gamma_\R \oplus
\Gamma_\R^*/L_\R$.  A second remark is that shifts of $\xi$ by
elements of $\Gamma \oplus \Gamma^*$ do not affect the boundary state
$\lVert R, \xi \rAngle$. So we should consider $\xi = (\xi_X,
\xi_{X^*})$ as an element of the torus $X \times X^*/(L_\R/L)$. Let us
now investigate the geometrical significance of $\xi$. The simplest
case is a 0-brane ($R = -\Id$). In that case $L = \Gamma^*$, so we can
take $\xi = (\xi_X,0)$. We can interpret $\xi_X \in X$ as the position
of the 0-brane. To justify this interpretation note that for any $k
\in \Gamma^*$
\[
\begin{split}
  \normord{e^{2\pi i \langle k, X \rangle}} \lVert R, \xi \rAngle 
  &= \sum_{m \in \Gamma^*} e^{-2\pi i \langle m, \xi_X \rangle} 
     \normord{e^{2\pi i \langle k, X \rangle}} \lvert R, 0,m \rAngle 
   = \sum_{m \in \Gamma^*} e^{-2\pi i \langle m, \xi_X \rangle}
     \lvert R, 0,m+k \rAngle \\
  &= e^{2\pi i \langle k, \xi_X \rangle} \lVert R, \xi \rAngle.
\end{split}
\]
So $\lVert R, \xi \rAngle$ is an eigenvector of $\normord{e^{2\pi i
    \langle k, X \rangle}}$ with eigenvalue $e^{2\pi i \langle k,
  \xi_X \rangle}$. For arbitrary gluing matrices we can obtain a
similar expression, but the interpretation is not completely clear.
Let us write the bosonic field $X(z,\bar{z})$ as $X_L(z) +
X_R(\bar{z})$. The vertex operator corresponding to $\lvert w,m
\rangle$ can then be written in a slightly imprecise notation as
$\normord{e^{2 \pi i(\langle k_L,X_L(z) \rangle + \langle k_R,
    X_R(\bar{z}) \rangle)}}$, where $k_L$ and $k_R$ are defined in
terms of $(w,m) \in L$ (see \pref{eq:kLR}).  For $(w,m) \in L$ a
computation analogous to the one above shows
\[
  \normord{e^{2 \pi i(\langle k_L,X_L(z) \rangle 
    + \langle k_R, X_R(\bar{z}) \rangle)}} \lVert R, \xi \rAngle
  = e^{2\pi i q((w,m), \xi)} \lVert R, \xi \rAngle.
\]
Defining $\tilde{X}(z,\bar{z}) = X_L(z) - X_R(\bar{z})$, we can
rewrite the operator in the exponent as follows
\[
  \langle k_L,X_L(z) \rangle + \langle k_R, X_R(\bar{z}) \rangle 
  = q \biggl ( \begin{twomatrix} w\\m \end{twomatrix}, 
   \begin{twomatrix} 
     X(z,\bar{z}) \\
     G \tilde{X}(z, \bar{z}) + B X(z, \bar{z})
   \end{twomatrix} \biggr ).
\]
So we can interpret the equation above as the statement that the
boundary state $\lVert R, \xi \rAngle$ is a joint eigenvector of the
vector of operators $(X(z,\bar{z}),G \tilde{X}(z, \bar{z}) + B X(z,
\bar{z}))$ with eigenvalues $\xi$ up to elements of $L_\R$ and up to
elements of the lattice $\Gamma \oplus \Gamma^*$. This seems to fit
the geometrical interpretation of next section quite well, but we do
not have such a clear interpretation as for the case $R = -\Id$.

As noted by Gaberdiel in~\cite{Gab02} we could generalise these
boundary states by taking $\xi$ in $\Gamma_\C^{\vphantom{*}} \oplus
\Gamma_\C^*$. This may cause problems with unitarity. A different
generalisation is to introduce Chan-Paton factors. In modern
terminology Chan-Paton factors correspond to multiple D-branes stacked
on top of each other.  For $r$ stacked D-branes Chan-Paton factors can
be implemented by replacing the state space $V$ by $V \otimes
M_r(\C)$. The fields also take values in $M_r(\C)$ and the expectation
values are changed to include a trace
\[
  \langle \phi_1 \dots \phi_k \rangle 
  := \Tr(\langle \phi_1 \dots \phi_k \rangle )
   = \sum_{i_1,\dots,i_k=1}^r
       \langle \phi_{1,i_1i_2} \dots \phi_{k,i_ki_1} \rangle.
\]
Because the state space is tensored with $M_r(\C)$, the same should be
true for the boundary states. Such boundary states can easily be
constructed by replacing $\xi$ in \pref{eq:bdystate} by a matrix
valued vector $\Xi \in (\Gamma_\R \oplus \Gamma_\R^*) \otimes
M_r(\C)$. The leads to the following more general definition of
boundary states
\begin{equation}
\label{eq:bdystateXi}
  \lVert R, \Xi \rAngle = \sum_{(w,m) \in L} e^{-2 \pi i q((w,m), \Xi)}
  \lvert R, w,m \rAngle.
\end{equation}
We will impose two restrictions on $\Xi \in (\Gamma \oplus
\Gamma^*) \otimes M_r(\C)$. The components $\Xi_i \in M_k(\C)$ of
$\Xi$ ($i=1, \dots, 2m$) should commute and all their eigenvalues
should be real. In addition we have to identify $\Xi$'s that lead to
the same boundary state. This leads to the following definition of the
space of \emph{Chan-Paton matrices}.
\begin{dfn}
\label{def:CP}
Let $K$ be in $\TM_{N=1}(\Gamma)$ and let $R$ be a gluing matrix, then
the space of rank $r$ Chan-Paton matrices is defined as
\[
  \mathrm{CP}(K, R, r) :=
  \biggl \{ \Xi \in (\Gamma_\R \oplus \Gamma_\R^*) \otimes M_r(\C)
  \biggm | \negthickspace
  \begin{array}{l}
    \text{\upshape all $\Xi_i$ ($i=1,\dots,2\rk \Gamma$) commute} \\
    \text{\upshape and have real eigenvalues}
  \end{array} \biggr \}\bigg /\negthickspace\sim,
\]
where $\sim$ is the equivalence relation defined by
\[
  \Xi \sim \Xi' \quad \Leftrightarrow \quad \exists \,\xi \in \Gamma
  \oplus \Gamma^* \:\forall (w,m) \in L_{R,K}: q((w,m), \Xi' - \Xi) =
  q((w,m), \xi)\Id_r.
\]
\end{dfn}
Using the definition of a boundary state it is easy to check that
$\lVert R,\Xi \rAngle = \lVert R,\Xi' \rAngle$ when $\Xi \sim
\Xi'$. This equivalence relation also has the following more explicit
description
\begin{equation}
\label{eq:Xiequiv}
  \Xi \sim \Xi' \quad \Leftrightarrow \quad \exists \,\xi \in \Gamma
  \oplus \Gamma^* \exists \, Y \in L_{R,K} \otimes M_r(\C): 
    \Xi' = \Xi + Y + \xi \Id_r.
\end{equation}
For $r=1$ this space coincides with the space $X \times
X^*/((L_{R,K} \otimes\R)/L_{R,K})$ we found before.

When the D-brane is a vector bundle $E$ (i.e., the underlying
submanifold is all of $X$), then the introduction of Chan-Paton
matrices should correspond to replacing $E$ by $E \otimes F$, where
$F$ is a flat vector bundle of rank $r$ with monodromies given by
$\Xi_i$ ($i=m+1, \dots 2m$). We will discuss this in more detail in
the next section.

Combining the gluing matrix with the Chan-Paton matrices leads to the
following definition of the set of D-branes.
\begin{dfn}
  The set of rank $r$ affine $N=1$ D-branes on a torus $X =
  \Gamma_\R/\Gamma$ corresponding to a point $K \in \TM_{N=1}$ in the
  Teichmüller space is defined as follows
\[
  C_K^r(X) := \{ (R,\Xi) \mid
    \text{$R: \Gamma_\R \rightarrow \Gamma_R$ satisfies 
      conditions~\ref{cond:ortho}, \ref{cond:rat1},
    and~\ref{cond:rat2}, $\Xi \in \mathrm{CP}(K, R, r)$} \}.
\]
The set $C_K(X)$ of all D-branes is the union $\cup_{r>0} C_K^r(X)$.
\end{dfn}
The transformation behaviour of these boundary states under $N=1$
isomorphisms is easy to determine. 
\begin{thm}
\label{thm:phig}
  An isomorphism $f_g$ of $N=1$ SCFTs corresponding to a lattice
  isomorphism $g: (\Gamma \oplus \Gamma^*,q) \rightarrow (\Gamma'
  \oplus {\Gamma'}^*, q')$ induces a bijective map $\phi_g: C_K^r(X)
  \rightarrow C_{K'}^r(X')$ defined by
  \[
    \phi_g(R,\Xi) = (R', g \Xi),
  \]
  with $R'$ as in \sref{tab:trules}. This map is compatible with $f_g$
  in the sense that
  \[
    \lVert R', g\Xi \rAngle = f_g \lVert R, \Xi \rAngle.
  \]
\end{thm}
\begin{proof}
  As already observed above, \sref{thm:LR} and \sref{prop:transLR}
  together imply that $R'$ satisfies the conditions~\ref{cond:ortho},
  \ref{cond:rat1}, and~\ref{cond:rat2} in terms of the metric and
  B-field corresponding to $K'$. To see that $\Xi' = g \Xi \in
  \mathrm{CP}(K', R', r)$, note that the lattice isomorphism $g$ only
  acts on the $\Gamma_\R \oplus \Gamma_\R^*$ part of the tensor
  product $(\Gamma_\R \oplus \Gamma_\R^*) \otimes M_r(\C)$. Therefore
  the components $\Xi'_i \in M_r(\C)$ of $\Xi'$ are linear
  combinations with real (even integral) coefficients of the
  components of $\Xi$.  This shows that they again commute and have
  real eigenvalues.  Because $g$ is a lattice isomorphism $(\Gamma
  \oplus \Gamma^*,q) \rightarrow (\Gamma' \oplus {\Gamma'}^*, q')$ and
  $L_{R',K'} = g(L_{R,K})$, mapping $\Xi$ to $g \Xi$ is compatible with the
  equivalence relation $\sim$ and $\Xi'$ is a well defined element of
  $\mathrm{CP}(K', R', r)$. Combining all this it follows that
  $(R',\Xi') \in C_{K'}^r(X')$. Bijectivity follows from the
  functoriality property $\phi_{g_1} \circ \phi_{g_2} = \phi_{g_1
    \circ g_2}$, which is easily verified.

To see that $\phi_g$ is compatible with the map $f_g$ on the boundary
states, we first compute $f_g$ on Ishibashi states
\[
\begin{split}
  f_g \lvert R,w,m \rAngle &= \exp \bigl (-\sum_{\ell=1}^\infty
  \frac{1}{\ell} G(f_g \alpha_{-\ell} f_g^{-1}, R f_g
  \bar{\alpha}_{-\ell} f_g^{-1} ) \bigr )
  f_g \lvert w,m \rangle \\
  &= \exp \bigl (-\sum_{\ell=1}^\infty
  \frac{1}{\ell} G((a+bE)^{-1} \alpha'_{-\ell}, R (a-bE^t)^{-1}
  \bar{\alpha}'_{-\ell}) \bigr ) \lvert g(w,m) \rangle\\
  &= \lvert R',g(w,m) \rAngle.
\end{split}
\]
Here we started with the definition of the Ishibashi states in
\pref{eq:Ishi} and then used the transformation of the
$\alpha_{-\ell}$'s and $\bar{\alpha}_{-\ell}$'s from \pref{eq:fggens}
and finally used the expressions for $G'$ and $R'$ from
\pref{tab:trules}. Combining this with the definition
\pref{eq:bdystateXi} of boundary states, we can easily verify the
required compatibility.
\end{proof}

\subsection{Geometrical interpretation}
In this section we will try to translate the algebraic description of
D-branes in terms of conformal field theory into more geometrical
terms.  We have alluded to the geometrical description of D-branes
several times as a motivation for parts of our algebraic description.
A first guess would be to represent a D-brane in a target space $X$ as
a triple $(S,E,\nabla)$ of a submanifold $S \subset X$ and a vector
bundle $E$ on $S$ with connection $\nabla$. However, for D-branes of
positive codimension and rank $r>1$ this description ignores part of
the information contained in the Chan-Paton matrices $\Xi \in
\mathrm{CP}(K, R, r)$ introduced in the previous section. The
description proposed here solves this problem and seems to fit
everything known about the geometrical description of D-branes.

As we will see in the next section, some D-branes are expected to have
a description as coherent sheaves. That means that they should be
modules over the structure sheaf $\mathcal{O}_X$ (see the description
of skyscraper sheaves in~\cite{PZ98}). For general D-branes we are not
in a holomorphic context, but we could try to work with modules over
$C^\infty(X)$ instead. Note that because we are in a smooth context,
it is not necessary to use sheaves. According to the Serre-Swan
theorem smooth vector bundles on a compact smooth manifold $X$ can be
identified with projective modules over $C^\infty(X)$. The projective
module corresponding to a vector bundle $E$ is its space of smooth
global sections $\Gamma(E)$. Using $\Omega^k(E) := \Omega^k(X) \otimes
\Gamma(E)$, we can define a connection as a map $\nabla: \Omega^k(E)
\rightarrow \Omega^{k+1}(E)$ satisfying the Leibniz property
$\nabla(fs) = \dop f \otimes s + f \nabla s$. To describe a D-brane we
drop the condition that the module be projective.
\begin{dfn}[Geometrical D-brane]
  A geometrical D-brane on a manifold $X$ is a pair $(M,\nabla)$
  consisting of a module $M$ over $C^\infty(X)$ with a connection
  $\nabla: \Omega^k(M) \rightarrow \Omega^{k+1}(M)$ satisfying the
  Leibniz rule. Here $\Omega^k(M)$ is defined as $\Omega^k(X)
  \otimes_{C^\infty(X)} M$.
\end{dfn}
It is likely that ultimately some replacement for the projectivity
will be necessary. However, we will only consider concrete examples
corresponding to the affine D-branes discussed above and those
D-branes should have whatever additional property will be required.

To construct explicitly the module associated to an affine D-brane,
let us first recall the definition of a vector bundle on a torus in
terms of multipliers.
\begin{dfn}
\label{def:mult}
Let $\Lambda$ be a lattice, $F$ an antisymmetric map $\Lambda \otimes
\Lambda \rightarrow \Z$, and $\alpha_0$ and $\beta_0$ matrix valued
vectors in $\Lambda^*_\R \otimes M_r(\C)$. Suppose that all components
of $\alpha_0$ and $\beta_0$ commute. Then we define a
\emph{multiplier} $e(x,\lambda) := e^{2\pi i \langle F x + \beta_0,
  \lambda \rangle}$ and a \emph{connection $1$-form} $\alpha := 2 \pi
i \langle F x + \alpha_0, \dop x \rangle$. These define a vector
bundle $E_{F,\alpha_0,\beta_0}$ on the torus $\Lambda_\R/\Lambda$ with
\[
  \Gamma(E_{F, \alpha_0,\beta_0}) := \{ s: \Lambda_\R \rightarrow \C^r \mid
  s(x+\lambda) = e(x,\lambda)~\text{for all $\lambda \in \Lambda$} \}
\]
as its space of sections. The connection $\nabla$ on this vector
bundle is defined by $\nabla s := \dop s + \alpha s$.
\end{dfn}
Note that using both $\alpha_0$ and $\beta_0$ is redundant, because
$E_{F,\alpha_0,\beta_0}$ is isomorphic to $E_{F,\alpha_0 +
  \beta_0,0}$. These bundles can be written as a tensor product $L
\otimes E'$, where $L$ is a line bundle with first Chern class given by
$F$ and $E'=E_{0,\alpha_0,\beta_0}$ is a flat vector bundle with rank
$r$.

Let $(R, \Xi)$ be an element of $C_K^r(X)$. As we saw above the gluing
matrix $R$ defines subspaces $V$ and $W$ of $\Gamma_\R$, such that
$\Gamma \cap W$ has maximal rank and an antisymmetric map $F: (\Gamma
\cap W) \otimes (\Gamma \cap W) \rightarrow \Q$. If we suppose that
$F$ is in fact integral, then we can apply this definition to $\Lambda
= \Gamma \cap W$ and $F$ to obtain a vector bundle on the subtorus
$W/(\Gamma \cap W)$. This is in fact the starting point of our
definition. Two issues complicates things a bit: we have to
account for $\Xi$ and we need a module over $C^\infty(X)$ instead of
over $C^\infty(W/(\Gamma \cap W)$. The following definition solves
both problems.
\begin{dfn}
  Let $(R, \Xi)$ be an element of $C_K^r(X)$, then we define the
  associated geometrical D-brane $(M,\nabla)$ as follows. Write
  $\Gamma_\R = V \oplus W$ and $\Gamma_\R^* = V^* \oplus W^*$ as
  above.  Suppose that $F: (\Gamma \cap W) \otimes (\Gamma \cap W)
  \rightarrow \Q$ is integral. Then we can apply \sref{def:mult} to
  the lattice $\Gamma \cap W$ to obtain a module $M:=
  \Gamma(E_{F,\alpha_0, \beta_0})$, where $\alpha_0 = \Xi_{W^*}$ and
  $\beta_0 = F \Xi_W$. The connection $\nabla$ is the one from
  \sref{def:mult}. To make $M$ into an $C^\infty(X)$ module we define
  \[
    f \cdot s(x) = (e^{\langle \frac{\partial}{\partial x}, 
      \Xi_X \rangle}f)(x)s(x).
  \]
\end{dfn}
A point that deserves some further explanation is the definition of
the module structure. To understand the notation write the components
of $\Xi_X$ in Jordan normal form $\Xi_X = \xi_X + N$, where $\xi_X \in
\Gamma_\R$ and $N$ is a vector of nilpotent matrices. Here it is
essential that we assumed the eigenvalues to be real (see
\sref{def:CP}). For simplicity we also assumed that there is just one
Jordan block. In this notation the action can be written as follows
\[
  f \cdot s(x) = (e^{\langle \frac{\partial}{\partial x}, N \rangle}
  f)(x + \xi_X) s(x) = \Bigl (\sum_{\alpha} \tfrac{1}{\alpha!}
  \bigl ( \tfrac{\partial}{\partial x} \bigr )^\alpha
  f(x + \xi_X) N^\alpha \Bigr ) s(x).
\]
The sum over the multi indices $\alpha$ is finite because the
component matrices of $N$ are nilpotent. Here it is important to note
that the differentiation and translation are not restricted to
directions parallel to $W$. The restriction to $x$ in $W$, which is
necessary to obtain an element of $M$, takes place afterwards.  From
this expression it is clear that we can consider $\xi_X$ to define the
position of the D-brane. This allows us to define the \emph{support}
of the D-brane $(R, \Xi)$ as
\begin{equation}
\label{eq:suppRXi}
  \supp(R, \Xi) = \xi_X + W/(\Gamma \cap W) \subset X.
\end{equation}
If there is more than one Jordan block, there are several
D-branes at different positions corresponding to the eigenvalues.

Note that in the expression above we describe the function $f$ on $X$
by a $\Gamma$-periodic function on $\Gamma_\R$. Together with the fact
that all components of $\Xi$ commute this ensures that $f
\cdot s(x + \lambda) = e(x,\lambda) f \cdot s(x)$.  The final thing to
check is the Leibniz rule. This is a straightforward computation.

Because of the assumption that $F$ be integral, instead of just
rational as required by Condition~\ref{cond:rat2}, this definition is not
completely general. The necessary generalisation is more or less
clear, but the necessary modifications would make the discussion less
transparent. Therefore, we will briefly sketch the solution, but then
return to the integral case for rest of this section.

The key idea is that we can replace the lattice $\Gamma_W := \Gamma
\cap W$ by a coarser lattice $\Gamma_W' \subset \Gamma_W$ such that
$F: \Gamma_W' \otimes \Gamma_W' \rightarrow \R$ is integral. In this
way we obtain a vector bundle $E'$ on $W/\Gamma_W'$. Using the isogeny
$i: W/\Gamma_W' \rightarrow W/\Gamma_W$ we can define a vector bundle
$E := i_*E'$ on $W/\Gamma_W$. The rank of this bundle is equal to the
degree of the isogeny times the rank of $E'$. Note that the
sublattice $\Gamma_W'$ is not unique, but if it is chosen such that
the degree of the isogeny is minimal, the vector bundle $E$ on
$W/\Gamma_W$ should be unique up to isomorphism. The module $M$ can
then be defined as the space of sections of the vector bundle $E$.

Let us now return to integral $F$. So far we have ignored a potential
problem with the definition above, namely the fact that $\Xi$ is only
determined up to the equivalence relation from \sref{def:CP}. The
following lemma shows that different choices lead to isomorphic
geometrical D-branes.
\begin{lemma}
Let $(M,\nabla)$ and $(M', \nabla')$ be geometrical D-branes
corresponding to the same gluing matrix and different, but equivalent
Chan-Paton matrices $\Xi$ and $\Xi'$, then they are isomorphic.
\end{lemma}
\begin{proof}
  We use the alternative description \pref{eq:Xiequiv} of the
  equivalence relation, i.e., we write $\Xi' = \Xi + Y + \xi \Id_r$,
  where $Y \in L_{R,K} \otimes M_r(\C)$ and $\xi \in \Gamma \oplus
  \Gamma^*$. Let us define a map $\phi: M \rightarrow M'$ by
\[
  \phi(s(x)) = (e^{\langle \frac{\partial}{\partial x}, \Xi'_W - \xi_W
  \rangle}) (e^{-\langle \frac{\partial}{\partial x}, \Xi_W
  \rangle}) s(x)
\]
We should check that $\phi(s)$ is indeed in $M'$ and that the map
$\phi$ is compatible with all the structures present. Because $s$ is
in $M$, we know that
\[
  s(x+\lambda) = e(x,\lambda) s(x) = e^{2 \pi i\langle F(x + \Xi_W),
  \lambda \rangle} s(x).
\]
For our computations the following identity will be very useful
\begin{equation}
\label{eq:transcomm}
  e^{\langle \frac{\partial}{\partial x}, A\rangle}(g(x)h(x)) =
  (e^{\langle \frac{\partial}{\partial x}, A\rangle} g(x))
  (e^{\langle \frac{\partial}{\partial x}, A\rangle} h(x)).
\end{equation}
This holds whenever all components $A_i$ of $A$ commute with $g(x)$.
When applying $\phi$ to the expression for $s(x+\lambda)$, we use this
twice to find
\[
  \phi(s(x+\lambda)) = (e^{\langle \frac{\partial}{\partial x},
    \Xi'_W -\xi_W \rangle}) \bigl (e^{2 \pi i\langle Fx, \lambda \rangle} 
  (e^{-\langle \frac{\partial}{\partial x}, \Xi_W \rangle} s(x))\bigr )
  = e^{2 \pi i\langle F(x+\Xi'_W), \lambda \rangle} \phi(s(x)).
\]
In the first step we could apply \pref{eq:transcomm}, because all
components of $\Xi$ commute. That turns the first factor into a
scalar, so that also in the second step there are no problems with non
commuting matrices. We omitted $\xi_W$ in the final expression,
because $\langle F \xi_W, \lambda \rangle$ is integral because of our
assumption that $F$ is integral. This shows that $\phi(s)$ is in $M'$.
The computation for the connection is slightly more complex
\[
\begin{split}
  \phi(\nabla s(x)) &= \phi(\dop s(x) + 2\pi i \langle F x + \Xi_{W^*},
  \dop x \rangle s(x)) \\
  &= (e^{\langle \frac{\partial}{\partial x}, \Xi'_W - \xi_W \rangle})
  (\dop (e^{-\langle \frac{\partial}{\partial x}, \Xi_W \rangle} s(x))\\
  & \qquad + 2\pi i \langle Fx-F\Xi_W  + \Xi_{W^*}, \dop x \rangle 
    (e^{-\langle \frac{\partial}{\partial x}, \Xi_W \rangle} s(x)))\\
  &= (e^{\langle \frac{\partial}{\partial x}, \Xi'_W - \xi_W \rangle})
  (\dop (e^{-\langle \frac{\partial}{\partial x}, \Xi_W \rangle} s(x))\\ 
  &\qquad + 2\pi i \langle Fx-F\Xi'_W+\Xi'_{W^*}+F\xi_W -\xi_{W^*}, 
    \dop x \rangle 
    (e^{-\langle \frac{\partial}{\partial x}, \Xi_W \rangle} s(x)))\\
  &= \dop (\phi(s(x))) + 2\pi i \langle Fx+\Xi'_{W^*} - \xi_{W^*},
    \dop x \rangle \phi(s(x)).
\end{split}
\]
Here we used again $\Xi' = \Xi + Y + \xi \Id_r$. Because $Y$ is in
$L_{R,K} \otimes M_r(\C)$, we know that $FY_W = Y_{W^*}$ (see
\pref{eq:LRW}).  Now recall that $\xi$ is an element of $\Gamma \oplus
\Gamma^*$. This means that $\langle \xi_{W^*}, \dop x \rangle$ defines
an integral 1-form on $W/(\Gamma \cap W)$, so $\phi$ takes the
connection on $M$ to a connection on $M'$ equivalent to $\nabla'$.
Finally we check the compatibility of $\phi$ with the module structure
\[
\begin{split}
  \phi(f \cdot s(x)) &= (e^{\langle \frac{\partial}{\partial x},
    \Xi'_W - \xi_W \rangle}) (e^{-\langle \frac{\partial}{\partial x}, 
    \Xi_W \rangle}) ((e^{\langle \frac{\partial}{\partial x}, 
    \Xi_X \rangle}f)(x) s(x)) \\
  &= (e^{\langle \frac{\partial}{\partial x}, \Xi'_W - \xi_W \rangle})
    ((e^{\langle \frac{\partial}{\partial x}, \Xi_V \rangle}f)(x) 
    (e^{-\langle \frac{\partial}{\partial x}, \Xi_W \rangle}) s(x))\\
  &= (e^{\langle \frac{\partial}{\partial x}, \Xi'_X - \xi_X \rangle}
     f)(x) \phi(s(x)) = f \cdot \phi(s(x)).
\end{split}
\]
Because $Y_V$ vanishes (see \pref{eq:LRV}), we have $\Xi_V = \Xi'_V -
\xi_V$. In the last step we use that $f$ is periodic, so $f(x-\xi_X) =
f(x)$.
\end{proof}
The proof of this lemma heavily uses the properties of the Chan-Paton
matrices we defined in the previous section. This lends support to our
definition of geometrical D-branes.  On the other hand this
geometrical definition sheds some light on the interpretation of $\Xi
\in \mathrm{CP}_K^r(X)$.

Finally note that it is expected that the geometry becomes non
commutative when $B$ does not vanish. However, at this level of detail
that seems to have no visible effects. This may be related to our
interpretation of $F = \mathcal{F} + \tilde{B}$ as the curvature matrix.

\subsection{D-brane categories}
This section is rather speculative. We will rephrase some of the
results we obtained so far in the language of D-brane categories and
use them to motivate a number of conjectures.

All D-branes of a given type together define a D-brane category. The
objects of such a category are the D-branes and the morphisms
correspond to strings stretching between a pair of D-branes.  So let
us see if we can construct the category $\mathcal{C}_K(X)$ of affine
D-branes on a torus $X$. The objects are the affine D-branes discussed
above. Often one also requires direct sums of D-branes to define an
object in the category. This corresponds to the fact that one can put
several D-branes together. Of course, such a configuration will be
unstable, but we will mostly ignore questions about stability.
Starting with a certain class of D-branes, we can define the objects
to be formal direct sums of these D-branes. In this way the class of
objects will automatically be closed under direct sums. A subtlety is
that direct sums of D-branes corresponding to the same gluing matrix
are already included, because we can simply take the direct sum of the
Chan-Paton matrices.  A D-brane category depends on where we are in
the moduli space of the SCFT\@. This can be seen in our example from
the fact that the set of affine D-branes $C_K(X)$ depends on $K \in
\TM_{N=1}(\Gamma)$.

The really problematic part in the construction of a D-brane category
is to define the morphisms and the composition of morphisms. In
general a D-brane category is expected to be a graded linear category,
i.e., the spaces of morphisms are graded vector spaces and the
composition is expected to be linear and compatible with the grading.
To construct the space of morphisms, one should analyse the conformal
field theory on the upper half plane with one D-brane on the positive
real axis and another one on the negative real axis. The Hilbert space
of this boundary conformal field theory (BCFT) is the space of
morphisms between the two D-branes and it is graded by fermion degree.
Elements in this Hilbert space can be identified with vertex operators
on the boundary.  Unfortunately the product of two fields in a CFT is
in general not defined. So it seems that our whole construction breaks
down at this point. However, by restricting the class of D-branes we
consider, we also restrict the class of vertex operators under
consideration. In addition we can replace the Hilbert space of the
boundary conformal field theory by some subspace. Together these
restrictions on the vertex operators that we consider may be enough to
guarantee that for those vertex operators a sensible product can be
defined.  Below we will discuss two cases where a construction along
these lines seems to work.

An alternative way out is to drop the requirement that a D-brane
category be a true category. Associating D-branes with objects and the
Hilbert spaces in BCFT with $\Hom$-spaces, seems to work quite well,
but the composition should be replaced with some other structure.
This is not as absurd as it may seem. After all an $A_\infty$-category
is also not a category in the strict sense. For such D-brane
categories it should still be possible to define notions like functor,
equivalence etc.  To find a replacement for the composition we can
probably use some of the structure defined by a BCFT (expectation
values, OPEs etc.). The real question is if we can obtain a useful
structure that can be described mathematically. We will not try to
make this precise here, but let us assume for the moment that we have
defined an appropriate D-brane category $\mathcal{C}_K(X)$. Because
$\mathcal{C}_K(X)$ is defined in terms of conformal field theory, the
whole construction should be compatible with $N=1$ isomorphisms. For
tori that means that each lattice isomorphism $g: (\Gamma \oplus
\Gamma^*, q) \rightarrow (\Gamma' \oplus {\Gamma'}^*, q')$ defines a
functor $\Phi_g: \mathcal{C}_K(X) \rightarrow \mathcal{C}_{K'}(X')$.
On objects this functor coincides with the map $\phi_g$ from
\sref{thm:phig}. Because of the corresponding property for $N=1$
isomorphisms, we also expect that $\Phi_{g_2} \circ \Phi_{g_1} =
\Phi_{g_2 \circ g_1}$. However, as we will discuss below, this is
probably too optimistic and we should be content with both sides being
isomorphic as functors (of D-brane categories).

\begin{table}
\caption{Boundary conditions}
\label{tab:ABbdy}
\begin{center}
\begin{tabular}{lxxr}
\toprule
&\multicolumn{2}{c}{Algebraic} & \multicolumn{3}{c}{Geometric} \\
\midrule
A-type & J(z) &= -\bar{J}(\bar{z})&
R^*\omega_1 &= -\omega_2 & (A1)\\
($\epsilon = -1$) &   G_{\pm}(z) &= \eta \Gbar_{\mp}(\bar{z})&
& & \\
&  e^{i\phi(z)} &= \eta^n e^{i\alpha} e^{-i\bar{\phi}(\bar{z})}&
R^*\Omega_1 &= e^{i\alpha} \bar{\Omega}_2 & (A2)\\
\midrule
B-type & J(z) &= \bar{J}(\bar{z})&
R^*\omega_1 &= \omega_2 & (B1)\\
($\epsilon = 1$) &   G_{\pm}(z) &= \eta \Gbar_{\pm}(\bar{z})&
& & \\
&  e^{i\phi(z)} &= \eta^n e^{i\beta} e^{i\bar{\phi}(\bar{z})}&
R^*\Omega_1 &= e^{i\beta} \Omega_2 & (B2)\\
\bottomrule
\end{tabular}
\end{center}
\end{table}

Let us now return to the first solution, restricting the class of
D-branes and morphisms. This allows us to stick to ordinary categories or
well known generalisations thereof such as $A_\infty$-categories. If
we choose an $N=2$ structure, we can define two classes of BPS
D-branes, the so called A- and B-branes. These D-branes can also be
considered as D-branes in a topological string theory obtained from
the original theory by the A- or B-twist respectively. Because
topological string theory is much simpler than ordinary string theory,
that gives a physical explanation why the behaviour of A- and B-branes
is so much nicer. The A- and B-type boundary conditions can be found
in \sref{tab:ABbdy}.  Algebraically these conditions can be formulated
in terms of the fields defining the $N=2$ structure. Note that this
means that to define A- and B-branes it is necessary to use the second
definition of an $N=2$ algebra discussed in \pref{sec:gensit}.  Using
the definition \pref{eq:Ntwogens} of these fields on a torus and the
boundary conditions \pref{eq:R}, these algebraic conditions can be
translated into geometrical terms.

Following~\cite{OOY96} we also use the spectral flow operators
\[
  e^{i\phi} = \Omega_1(\psi, \dots, \psi),
\]
and similarly for the right-moving sector. Here $\Omega_i$ is the
holomorphic volume form corresponding to the complex structure $j_i$.
In the conditions involving these operators there are real parameters
$\alpha$ and $\beta$. The boundary conditions are that there exists an
$\alpha \in \R$ or a $\beta \in \R$ such that the equation holds.

Actually for affine D-branes the conditions (A2) and (B2) are implied
by (A1) and (B1) respectively. This can be seen as follows. Recall
that $R^t G R = G$ (see Condition~\ref{cond:ortho}) and that $\omega_i
= -G j_i$. Let us do B-type boundary conditions first. We can write
(B1) as $R^t \omega_1 R = -\omega_2$. Then it follows that $j_1 R = R
j_2$, which implies that $R^* \Omega_1 = \lambda \Omega_2$ for some
$\lambda \in \C$ (a holomorphic 1-form $\alpha$ satisfies $\alpha
\circ j_1 = i \alpha$, so $R^*\alpha \circ j_2 = \alpha \circ R \circ
j_2 = \alpha \circ j_1 \circ R = i \alpha \circ R = i R^*\alpha$).
Using the normalisation $\Omega_i \wedge \bar{\Omega}_i = \frac{1}{n!}
\omega_i^n$ it follows that $|\lambda| =1$, so we can write $\lambda =
e^{i\beta}$ for some $\beta \in \R$. Pointwise this argument is always
valid (also on more general manifolds and for more general D-branes).
However, the essential point is that $\beta$ should be constant. For
tori and affine D-branes that is automatic, but for more general
situations that is a highly nontrivial condition.

For A-type boundary conditions the argument is similar. Using (A1) we
find $j_1 R = - R j_2$. This implies $R^* \Omega_1 = \lambda
\bar{\Omega}_2$ for some $\lambda \in \C$. Using the normalisation of
$\omega_1$ and $\Omega_1$ it follows that $(-1)^n |\lambda|^2 \Omega_2
\wedge \bar{\Omega}_2 = (-1)^n \frac{1}{n!} \omega_2^n$. Comparing
with the normalisation of $\omega_2$ and $\Omega_2$ yields $|\lambda|
= 1$. Below we will discuss some hints that the conditions (A2) and
(B2) can be thought of as stability conditions. Apparently affine
D-branes are automatically stable, which explains why these conditions
are automatically satisfied in that case.

Using A- and B-type boundary conditions we can define two subsets of
$C_{IJ}(X)$, namely the set $C_{I,J}^A(X)$ of A-branes and the set
$C_{I,J}^B(X)$ of B-branes. Starting with these sets of D-branes we
hope to construct the corresponding D-brane categories
$\mathcal{C}^{\mathrm{A}}_{I,J}(X)$ and
$\mathcal{C}^{\mathrm{B}}_{I,J}(X)$. When $(I,J)$ is in the
geometrical part of the Teichmüller space, these categories should
also have a geometrical description, as was discussed in~\cite{OOY96}.
This is easiest for the B-branes, so let us discuss that case first.
\begin{conj}
\label{conj:BisDer}
  For $(I,J) \in \TMp_{N=2}^{\mathrm{geom}}$ the category of B-branes
  $\mathcal{C}^{\mathrm{B}}_{I,J}(X)$ is equivalent to the derived
  category of coherent sheaves $\mathcal{D}\mathrm{Coh}_{I,J}(X)$.
\end{conj}
To make this conjecture plausible, let us argue that an object
$(R,\Xi) \in C^\mathrm{B}_{I,J}(X)$ corresponds to a coherent sheaf on
$X$ (following~\cite{OOY96}). Because $(I,J) \in
\TMp_{N=2}^{\mathrm{geom}}(X)$, there is a single complex structure $j
= j_1 = j_2$ and a single Kähler form $\omega = \omega_1 = \omega_2 =
- Gj$ on $X$. So the equation $j_1 R = R j_2$ we found above
simplifies to $j R = R j$. This means that $R : \Gamma_\R \rightarrow
\Gamma_\R$ is a complex linear map if we regard $\Gamma_\R$ as a
complex vector space with complex structure defined by $j$. So the
$-1$ eigenspace $V \subset \Gamma_\R$ is a complex subspace of
$(\Gamma_\R, j)$. Because $j^t G j = G$ it follows that the orthogonal
complement $W$ is a complex subspace as well. Because the support
$\supp(R,\Xi)$ as defined in \pref{eq:suppRXi} is a shift of
$W/(\Gamma \cap W)$, it is a holomorphic submanifold of $X$. Using the
fact that $j^{-1} \tilde{R} j = \tilde{R}$ and $j^t \tilde{G} j =
\tilde{G}$ it follows that $\tilde{R} = (\tilde{G} - j^t \mathcal{F}
j)^{-1} (\tilde{G} + j^t \mathcal{F} j)$, so because of the uniqueness
of $\mathcal{F}$ we see that $j^t \mathcal{F} j = \mathcal{F}$. In
other words $\mathcal{F}$ is a $(1,1)$-form, so the vector bundle $E$
on $\supp(R,\Xi)$ is holomorphic. This argument only works when $B$ is
a $(1,1)$-form. If it is not, we have to twist the derived category
(see \cite{KO00})

As discussed above the condition (B2) is automatically fulfilled. A
physical argument that affine D-branes are likely to be stable is that
they have minimal volume (mass) in their cohomology class. This only
says something about the support, so it cannot be a complete argument.
Mathematically, one can check that the vector bundles we obtain are
precisely the semi-homogeneous vector bundles discussed by Mukai
in~\cite{Muk78}. Mukai also shows that semi-homogeneous vector bundles
are Gieseker semi-stable and Gieseker stable when they are
simple. This seems to confirm our suspicion that the condition (B2) is
related to stability.

Not every object of the derived category corresponds to an element of
$C^\mathrm{B}_{I,J}(X)$. However, Fukaya in~\cite{Fuk99} mentions a
conjecture by Mukai that any coherent sheaf on an abelian variety has
a resolution in terms of semi-homogeneous bundles. That would mean that
although we only have a direct description of semi-homogeneous vector
bundles on affine submanifolds, we can still `approximate' arbitrary
objects of the derived category in terms of them. In other words,
semi-homogeneous vector bundles generate the derived category.

Actually, this is not too different from our understanding on physical
side. We know how to describe D-branes corresponding to
semi-homogeneous vector bundles on affine submanifolds in conformal
field theory, but it is unclear how to describe more general sheaves.
Recent work in physics suggests that the resolutions from mathematics
also have a physical significance (see e.g.,~\cite{Dou00,KS02,KPS02}).

The corresponding conjecture for A-branes is the following.
\begin{conj}
\label{conj:AisFuk}
  For $(I,J) \in \TMp_{N=2}^{\mathrm{geom}}$ the category of A-branes
  $\mathcal{C}^{\mathrm{A}}_{I,J}(X)$ is equivalent to the derived
  Fukaya category $\mathcal{DF}_{I,J}(X)$.
\end{conj}
Let $(R,\Xi) \in C^A_{I,J}(X)$ be an A-brane and let us suppose that
$\mathcal{F} = 0$. This means that $W$ is the eigenspace for
eigenvalue $+1$. So if $v,w \in W$, then $\omega(v,w) = - \omega(Rv,
Rw) = -\omega(v,w)$. Therefore the restriction of $\omega$ to
$\supp(R,\Xi)$ vanishes. Similarly, we see that $\omega(v,w) = 0$ for
$v,w \in V$. Because $V \oplus W = \Gamma_\R$ it follows that $W$ has
dimension $n$, so $\supp(R,\Xi)$ is a Lagrangian submanifold of $X$.
At least when $B|_W = 0$, we also have $F = 0$, so the vector bundle
$E$ on $S$ is flat. If $v_1,\dots,v_n \in W$ then $\Omega(v_1, \dots,
v_n) = R^*\Omega(v_1, \dots, v_n) = e^{i\alpha} \bar{\Omega}(v_1,
\dots, v_n)$, so $\Im (e^{-i\alpha/2} \Omega)|_S = 0$. Therefore
$\supp(R,\Xi)$ is even special Lagrangian and we find back the
original definition of objects in the Fukaya category. The notion of
being special Lagrangian is closely related to $\Pi$-stability (see
\cite{DFR00,Dou00}).

However, this argument depends on the assumption $\mathcal{F}=0$.
In~\cite{KO01} Kapustin and Orlov showed that without this assumption
one obtains a more general class of D-branes, the so-called
coisotropic D-branes. So the Fukaya category should be a kind of
generalised Fukaya category including coisotropic D-branes. Another
problem used to be the construction of a derived version of the Fukaya
category, but that problem has now been solved thanks to the work of
Fukaya and collaborators (see~\cite{FOOO00}).

It is natural to expect that isomorphisms of $N=2$ SCFTs correspond to
isomorphisms of these categories, but for generalised $N=2$ morphisms
it is not quite clear what to expect. It may be helpful to return to
the putative general D-brane category $\mathcal{C}_K(X)$ for a moment.
To keep track of the $N=2$ structure we define $\mathcal{C}_{I,J}(X) =
\mathcal{C}_K(X)$, where $K=IJ$. The functor $\Phi_g$ induces functors
$\Phi_g^{(\epsilon_L,\epsilon_R)}: \mathcal{C}_{I,J}(X) \rightarrow
\mathcal{C}_{\mu_g^{(\epsilon_L,\epsilon_R)}(I,J)}$. The justification
for this notation is that we want to be able to consider
$\mathcal{C}_{I,J}^\mathrm{A}(X)$ and
$\mathcal{C}_{I,J}^\mathrm{B}(X)$ as subcategories of
$\mathcal{C}_{I,J}(X)$. The functor
$\Phi_g^{(\epsilon_L,\epsilon_R)}$ need not preserve these
subcategories, because their definition depends on $J(z)$ and
$\bar{J}(\bar{z})$. However, generalised $N=2$ morphisms are well
behaved on these fields (see \pref{tab:genmor}). Comparing to the
algebraic formulation of A- and B-type boundary conditions in
\sref{tab:ABbdy}, one can easily check that
$f_g^{(\epsilon_L,\epsilon_R)}$ interchanges A- and B-type boundary
conditions when $\epsilon_L \epsilon_R = -1$ and preserves them when
$\epsilon_L \epsilon_R = 1$.  So the corresponding functor
$\Phi_g^{(\epsilon_L,\epsilon_R)}$ should interchange the A- and
B-type subcategories for $\epsilon_L \epsilon_R = -1$ and preserve
them for $\epsilon_L \epsilon_R = 1$.

To make this more precise we can restrict the functor
$\Phi_g^{(\epsilon_L,\epsilon_R)}$ to either
$\mathcal{C}_{I,J}^\mathrm{A}(X)$ or
$\mathcal{C}_{I,J}^\mathrm{B}(X)$. The image is again one of these
subcategories. As these subcategories are much better defined, one may
hope to be able to make sense of these restrictions even if it is
difficult to make sense of $\Phi_g^{(\epsilon_L,\epsilon_R)}$ itself.
To formulate a precise conjecture, it is useful to label the boundary
condition with $\epsilon = -1$ for type A and $\epsilon = 1$ for type
B. The first geometrical condition in \sref{tab:ABbdy} can then be
written as $R^* \omega_2 = \epsilon \omega_1$. Using this notation we
can state the following conjecture.
\begin{conj}
\label{conj:genfunc}
  Let $g: (\Gamma \oplus \Gamma^*,q) \rightarrow (\Gamma' \oplus
  {\Gamma'}^*,q')$ be a lattice isomorphism and let $(\epsilon_L,
  \epsilon_R) = (\pm 1, \pm 1)$. Then for all $(I,J) \in
  \TM_{N=2}(\Gamma)$ and $\epsilon = \pm 1$ there exists a functor
  $\Phi_{g,\epsilon}^{(\epsilon_L, \epsilon_R)} :
  \mathcal{C}_{I,J}^{\epsilon_{\vphantom{L}}}(X) \rightarrow
  \mathcal{C}_{I',J'}^{\epsilon_L \epsilon_R \epsilon}(X')$, where
  $(I',J') = \mu_g^{(\epsilon_L, \epsilon_R)}(I,J)$. These functors
  satisfy
\[
  \Phi_{g_2,\epsilon_{L,1} \epsilon_{R,1} \epsilon}^{(\epsilon_{L,2},
    \epsilon_{R,2})} \circ
  \Phi_{g_1,\epsilon}^{(\epsilon_{L,1}, \epsilon_{R,1})} 
  \cong \Phi_{g_2 \circ g_1,\epsilon}^{(\epsilon_{L,2} \epsilon_{L,1}, 
    \epsilon_{R,2} \epsilon_{R,1})},
\]
where $\cong$ denotes isomorphism of functors.
\end{conj}
Here $\Phi_{g,\epsilon}^{(\epsilon_L, \epsilon_R)}$ can be thought of
as the restriction of $\Phi_g^{(\epsilon_L, \epsilon_R)}$ to
$\mathcal{C}_{I,J}^\mathrm{A}(X)$ for $\epsilon = -1$ or to
$\mathcal{C}_{I,J}^\mathrm{B}(X)$ for $\epsilon = 1$. The statement
about the composition of functors reflects \sref{prop:genmorfunc}.
Note that we have been careful and do not require equality, but just
an isomorphism of functors in the last property. On the level of
objects these functors are given by $\phi_g$ and we do have
equality. However, as we saw above, the sets of D-branes
$C^{\mathrm{A}}_{I,J}$ and $C^{\mathrm{B}}_{I,J}$ and the morphisms
between them, are really just a starting point. To actually construct
the D-brane categories one has to add formal direct sums, construct
the derived category etc. After all these constructions it is rather
likely that the best one can hope for is an isomorphism of functors.
Taking $g_2 = g_1^{-1}$ and $(\epsilon_{L,2}, \epsilon_{R,2}) =
(\epsilon_{L,1}, \epsilon_{R,1})$ it follows from this property that
the functors $\Phi_{g,\epsilon}^{(\epsilon_L, \epsilon_R)}$ are
equivalences.

For $\epsilon_L \epsilon_R = -1$ these functors define equivalences
between the category of A-branes on $X$ and the category of B-branes
on $X'$. When we are in the geometrical part of the moduli space we
can use \sref{conj:BisDer} and \sref{conj:AisFuk} and interpret these
categories as the derived Fukaya category and the derived category of
coherent sheaves. So we see that this conjecture is in fact a
generalisation of Kontsevich's homological mirror symmetry conjecture
applied to tori (see~\cite{Kon94}). From the discussion of T-duality
in \pref{sec:geomint} it follows that mirror symmetry in the
geometrically interesting cases can be described as T-duality in the
fibres of a torus fibration (the SYZ-description). On the other hand
if we fix a torus fibration and consider the mirror morphism
corresponding to T-duality in the fibres of this fibration, then for
certain values of the background field this mirror morphism will take
us to a non geometrical part of the moduli space. In that case
Kontsevich's conjecture breaks down, but our conjecture still predicts
an equivalence of categories. However, we can no longer interpret both
categories as the derived Fukaya category or the derived category of
coherent sheaves.

This description also points to a way of breaking down the proof of
Kontsevich's conjecture into several steps. One part would be proving
\sref{conj:AisFuk} and \sref{conj:BisDer}. That would reduce the proof
to \sref{conj:genfunc}. Here we at least have a solid understanding of
the map on the moduli and we also know the map on the objects for a
large class of objects. In fact, the class of D-brane we describe is a
generalisation of the class used by Polishchuk and Zaslow in their
proof of the homological mirror symmetry conjecture for 2-dimensional
tori. One can check that our description leads to exactly the same map
on objects. The advantage of our description is that we have a largely
systematic description of the map on D-branes.  The disadvantage is
that so far a description of the morphisms and the composition is
lacking.

\section{Conclusions}
In this paper we tried to clarify some issues concerning the moduli
spaces of tori and their D-brane categories. We emphasised the need to
clearly distinguish the various moduli spaces. Only then is it
possible to understand the relations between the moduli spaces. For
tori we described these moduli spaces in great detail. For $N=2$ this
led us to a geometrical description using two independent complex
structures. In itself this structure is general, but for tori it is
particularly interesting because there are so many complex structures
compatible with a given metric. Together with a precise description
of isomorphisms of $N=1$ and $N=2$ SCFTs this allowed us to explain
some issues with interpretation of mirror symmetry for general values
of the background fields.

Turning our attention to D-branes, we found that this description of
the moduli space interacts nicely with the conformal field theory
description of D-branes using gluing matrices.  Using the boundary
state formalism we studied the transformation of D-branes under
isomorphisms of the SCFT. Explicit formulae for the boundary states
motivated the introduction of extra data complementing the gluing
matrix and necessary to completely determine a D-brane.

The final sections were more speculative. This also takes us to
options for further research. Our preliminary geometrical description
of D-branes should be made more precise and also the categorical
interpretation of D-brane needs further study. The best opportunity
for progress on these topics seems to be an even more detailed
investigation of the relation between conformal field theory and
geometry, especially the relation between boundary conformal field
theory and D-brane categories. That should lead to a better
understanding of how the Fukaya category and the derived category of
coherent sheaves have to be modified to obtain a completely general
description of the categories of A- and B-branes. That in turn should
improve our understanding of the homological mirror symmetry
conjecture.

\acknowledgments 
I want to thank the Isaac Newton Institute for hospitality. My
discussions there with Katrin Wendland about the moduli space of K3
surfaces were an important initial motivation for this work. I also
want to thank her for email correspondence which helped to shape my
ideas about the various moduli spaces and their relations.

Finally, I want to thank Sarah Dean, Markus Rosellen, Emanuel
Scheidegger, and Duco van Straten for discussions about my work.

This work was supported by the DFG-Schwerpunkt Stringtheorie. 

\bibliographystyle{JHEP}
\bibliography{categories,dbranes,thesis1,thesis2}

\providecommand{\href}[2]{#2}\begingroup\raggedright\begin{thebibliography}{10}

\bibitem{KO00}
A.~Kapustin and D.~Orlov, {\it Vertex algebras, mirror symmetry, and
  {D}-branes: The case of complex tori},  {\em Commun. Math. Phys.} {\bf 233}
  (2003) 79--136, [\href{http://xxx.lanl.gov/abs/hep-th/0010293}{{\tt
  hep-th/0010293}}].

\bibitem{OOY96}
H.~Ooguri, Y.~Oz, and Z.~Yin, {\it {D}-branes on {C}alabi-{Y}au spaces and
  their mirrors},  {\em Nucl. Phys.} {\bf B477} (1996) 407--430,
  [\href{http://xxx.lanl.gov/abs/hep-th/9606112}{{\tt hep-th/9606112}}].

\bibitem{AM94}
P.~S. Aspinwall and D.~R. Morrison, {\it String theory on {K}3 surfaces},
  \href{http://xxx.lanl.gov/abs/hep-th/9404151}{{\tt hep-th/9404151}}.

\bibitem{Dij98}
R.~Dijkgraaf, {\it Instanton strings and hyperkähler geometry},  {\em Nucl.
  Phys.} {\bf B543} (1999) 545--571,
  [\href{http://xxx.lanl.gov/abs/hep-th/9810210}{{\tt hep-th/9810210}}].

\bibitem{NW99}
W.~Nahm and K.~Wendland, {\it A hiker's guide to {K}3: Aspects of {N} = (4,4)
  superconformal field theory with central charge c = 6},  {\em Commun. Math.
  Phys.} {\bf 216} (2001) 85--138,
  [\href{http://xxx.lanl.gov/abs/hep-th/9912067}{{\tt hep-th/9912067}}].

\bibitem{NW01}
W.~Nahm and K.~Wendland, {\it Mirror symmetry on {K}ummer type {K}3 surfaces},
  \href{http://xxx.lanl.gov/abs/hep-th/0106104}{{\tt hep-th/0106104}}.

\bibitem{Sei88}
N.~Seiberg, {\it Observations on the moduli space of superconformal field
  theories},  {\em Nucl. Phys.} {\bf B303} (1988) 286--304.

\bibitem{GPR94}
A.~Giveon, M.~Porrati, and E.~Rabinovici, {\it Target space duality in string
  theory},  {\em Phys. Rept.} {\bf 244} (1994) 77--202,
  [\href{http://xxx.lanl.gov/abs/hep-th/9401139}{{\tt hep-th/9401139}}].

\bibitem{Huy01}
D.~Huybrechts, ``Moduli spaces of tori and the {N}arain moduli space.''
  Available from
  \href{http://www.mi.uni-koeln.de/~huybrech/narain.ps}{\texttt{http://www.mi.%
uni-koeln.de/~huybrech/narain.ps}}, 2001.

\bibitem{SYZ}
A.~Strominger, S.-T. Yau, and E.~Zaslow, {\it Mirror symmetry is {T}-duality},
  {\em Nucl. Phys.} {\bf B479} (1996) 243--259,
  [\href{http://xxx.lanl.gov/abs/hep-th/9606040}{{\tt hep-th/9606040}}].

\bibitem{Enc00}
C.~v. Enckevort, {\em Mirror Symmetry and {T}-duality, Equivalence of {D}-brane
  categories}.
\newblock PhD thesis, University of Utrecht, 2000.
\newblock Available from
  \href{http://enriques.mathematik.uni-mainz.de/enckevort/index.html}{\texttt{%
http://enriques.mathematik.uni-mainz.de/enckevort/index.html}}.

\bibitem{Sch02}
V.~Schomerus, {\it Lectures on branes in curved backgrounds},  {\em Class.
  Quant. Grav.} {\bf 19} (2002) 5781--5847,
  [\href{http://xxx.lanl.gov/abs/hep-th/0212218}{{\tt hep-th/0212218}}].

\bibitem{RS98}
A.~Recknagel and V.~Schomerus, {\it Boundary deformation theory and moduli
  spaces of {D}-branes},  {\em Nucl. Phys.} {\bf B545} (1999) 233--282,
  [\href{http://xxx.lanl.gov/abs/hep-th/9811237}{{\tt hep-th/9811237}}].

\bibitem{RS97}
A.~Recknagel and V.~Schomerus, {\it {D}-branes in {G}epner models},  {\em Nucl.
  Phys.} {\bf B531} (1998) 185--225,
  [\href{http://xxx.lanl.gov/abs/hep-th/9712186}{{\tt hep-th/9712186}}].

\bibitem{Gab02}
M.~R. Gaberdiel, {\it {D}-branes from conformal field theory},  {\em Fortsch.
  Phys.} {\bf 50} (2002) 783--801,
  [\href{http://xxx.lanl.gov/abs/hep-th/0201113}{{\tt hep-th/0201113}}].

\bibitem{ALZ01}
C.~Albertsson, U.~Lindström, and M.~Zabzine, {\it {$N=1$} supersymmetric sigma
  model with boundaries, {I}},  {\em Commun. Math. Phys.} {\bf 233} (2003)
  403--421, [\href{http://xxx.lanl.gov/abs/hep-th/0111161}{{\tt
  hep-th/0111161}}].

\bibitem{ALZ02}
C.~Albertsson, U.~Lindström, and M.~Zabzine, {\it {$N=1$} supersymmetric sigma
  model with boundaries, {II}},
  \href{http://xxx.lanl.gov/abs/hep-th/0202069}{{\tt hep-th/0202069}}.

\bibitem{Gab00}
M.~R. Gaberdiel, {\it Lectures on non-{BPS} {D}irichlet branes},  {\em Class.
  Quant. Grav.} {\bf 17} (2000) 3483--3520,
  [\href{http://xxx.lanl.gov/abs/hep-th/0005029}{{\tt hep-th/0005029}}].

\bibitem{PZ98}
A.~Polishchuk and E.~Zaslow, {\it Categorical mirror symmetry: The elliptic
  curve},  {\em Adv. Theor. Math. Phys.} {\bf 2} (1998) 443--470,
  [\href{http://xxx.lanl.gov/abs/math.AG/9801119}{{\tt math.AG/9801119}}].

\bibitem{Muk78}
S.~Mukai, {\it Semi-homogeneous vector bundles on an abelian variety},  {\em J.
  Math. Kyoto Univ.} {\bf 18-2} (1978) 239--272.

\bibitem{Fuk99}
K.~Fukaya, ``Mirror symmetry of abelian variety and multi theta functions.''
  Available from
  \href{http://www.kusm.kyoto-u.ac.jp/\symbol{126}fukaya/fukaya.html}{http://w%
ww.kusm.kyoto-u.ac.jp/\symbol{126}fukaya/fukaya.html}, 1998.

\bibitem{Dou00}
M.~R. Douglas, {\it {D}-branes on {C}alabi-{Y}au manifolds},
  \href{http://xxx.lanl.gov/abs/math.AG/0009209}{{\tt math.AG/0009209}}.

\bibitem{KS02}
S.~Katz and E.~Sharpe, {\it D-branes, open string vertex operators, and {E}xt
  groups},  \href{http://xxx.lanl.gov/abs/hep-th/0208104}{{\tt
  hep-th/0208104}}.

\bibitem{KPS02}
S.~Katz, T.~Pantev, and E.~Sharpe, {\it D-branes, orbifolds, and {E}xt groups},
   \href{http://xxx.lanl.gov/abs/hep-th/0209241}{{\tt hep-th/0209241}}.

\bibitem{DFR00}
M.~R. Douglas, B.~Fiol, and C.~R{\"o}melsberger, {\it Stability and {BPS}
  branes},  \href{http://xxx.lanl.gov/abs/hep-th/0002037}{{\tt
  hep-th/0002037}}.

\bibitem{KO01}
A.~Kapustin and D.~Orlov, {\it Remarks on {A}-branes, mirror symmetry, and the
  {F}ukaya category},  \href{http://xxx.lanl.gov/abs/hep-th/0109098}{{\tt
  hep-th/0109098}}.

\bibitem{FOOO00}
K.~Fukaya, Y.-G. Oh, H.~Ohta, and K.~Ono, ``Lagrangian intersection {F}loer
  theory, anomaly and obstruction.'' Available from
  \href{http://www.kusm.kyoto-u.ac.jp/~fukaya/fukaya.html}{\texttt{http://www.%
kusm.kyoto-u.ac.jp/~fukaya/fukaya.html}}, 2000.

\bibitem{Kon94}
M.~Kontsevich, {\it Homological algebra of mirror symmetry},  in {\em
  Proceedings of the 1994 International Congress of Mathematicians},
  pp.~120--139, Birkh{\"a}user, 1995.
\newblock \href{http://xxx.lanl.gov/abs/alg-geom/9411018}{{\tt
  alg-geom/9411018}}.

\end{thebibliography}\endgroup
\end{document}